\documentclass[twocolumn,showpacs,superscriptaddress,preprintnumbers,amsmath,amssymb,epsfig,color,superscriptaddress]{revtex4}
\usepackage[latin1]{inputenc}
\usepackage[usenames]{color}
\usepackage[usenames]{color}
\usepackage{graphicx}
\def\Nb{$\bar{N}\;$}
\def\ab{$\bar{a}\;$}

\def\pb{$\bar{p}\;$}
\def\nb{$\bar{n}\;$}
\def\pbs{$\bar{p}s\;$}
\def\db{$\bar{d}\;$}
\def\dbs{$\bar{d}s\;$}
\def\tb{$\bar{t}\;$}
\def\hetb{$\bar{^3He}\;$}
\def\heqb{$\bar{^4He}\;$}
\def\het{$^3\!H\!e\,$}
\def\he4{$^4\!H\!e\,$}
\def\3b{$\bar{3}\;$}

\begin{document}
\preprint{Report LPSC/04-88}
\input epsf
\title{Flux of light antimatter nuclei near earth,\\
induced by Cosmic Rays in the Galaxy and in the atmosphere.}

\author{R. Duperray}
\author{B. Baret}
\affiliation{Laboratoire de Physique Subatomique et de Cosmologie, CNRS/IN2P3,
53 avenue des Martyrs, 38026 Grenoble-cedex, France} 
\author{D. Maurin} 
\affiliation{Service d'Astrophysique,
SAp CEA-Saclay, F-91191 Gif-sur-Yvette CEDEX, France}
\author{G. Boudoul}\altaffiliation[Permanent address:]{IPNL,Universit\'e Claude Bernard, 4 rue E. Fermi, 69622 Villeurbanne-cedex, France}
\author{A. Barrau}
\author{L. Derome}
\author{K. Protasov}
\author{M. Bu\'enerd}\email[corresponding author:] {buenerd@lpsc.in2p3.fr}
\affiliation{Laboratoire de Physique Subatomique et de Cosmologie, CNRS/IN2P3,
53 avenue des Martyrs, 38026 Grenoble-cedex, France} 

\date{\today}

\begin{abstract}
The fluxes of light antinuclei A$\leq 4$ induced near earth by Cosmic Ray interactions with the
interstellar matter in the Galaxy and with the Earth atmosphere are calculated in a phenomenological 
framework. The hadronic production cross section for antinucleons is based on a recent parametrization 
of a wide set of accelerator data. The production of light nuclei is calculated using coalescence
models. For the standard coalescence model, the coalescence radius is fitted to the available experimental 
data. The non annihilating inelastic scattering process for the antideuterons is discussed and taken 
into account for the first time via a more realistic procedure than used so far for antiprotons.
\end{abstract}

\maketitle


\section{Introduction}

Antiprotons in Cosmic Rays (CR) have been extensively studied both experimentally and
theoretically over the last few decades, with the general purpose of measuring their
flux and understanding their origin.
It is now generally agreed that the dominant part of the \pb CR spectrum is a secondary
flux originating from the hadronic production induced by CRs on the interstellar (IS) 
medium (ISM). This agreement is grounded on the ability of the calculations based on this
assumption to reproduce the data.
Recently some new prospects have been outlined, strengthening the motivations for the
study of the \pb flux \cite{DO01}, and extending the interest for CR antimatter to
other light antinuclei, antideuterons \db in particular, with the emergence of new
astrophysical issues. The \pb and \db production in neutralino annihilation has been
considered as a possible signature for the Dark Matter constituents in the universe 
\cite{DBDM}.
Antiprotons and antideuterons have also been considered as evaporation products of
primordial black holes (PBH) and their flux at earth calculated in the perspective of 
searching for a possible signature of the source in the experimental spectra 
\cite{DBPBH,PBPBH}.
A common feature of these studies is that the calculated fluxes are extremely small in
intensity with their spectra peaked at low momenta. This requires the secondary galactic 
contribution to be accurately known for a significant search of the former to be 
considered. 

Another motivation for a careful examination of the antimatter production in the
Galaxy 
is provided by the need of evaluating the galactic flux which will constitute a physical
background for the forthcoming new experiments to search for primordial antimatter in the 
universe \cite{AMS01,PAM,BESSP,BESDBA}. The evaluation can be performed using the CR flux
data from the latest generation of experiments \cite{HA04,BO97,AMS01}.

In the atmosphere, The CR induced \pb flux has been measured recently \cite{BESS2770} at 
mountain altitude, and some antinuclei production is naturally expected to originate from the 
same source. These atmospheric secondaries will generate another physical background in the 
future experiments. Although it could be separated in principle from the galactic 
flux on dynamic and kinematic grounds, a soundly based knowledge of this background is 
required. Recently the phenomenology of the particle production induced by CRs in the 
atmosphere has been thoroughly investigated with the purpose of accounting for the large 
amount of new data available from recent ballon and satellite measurements 
\cite{DE01,HU03}. For the \pb flux, it has been shown in these works that this flux 
can be accounted for in a conceptually simple, albeit numerically elaborated, framework 
\cite{DE01,HU03}. These calculations are extended here to the case of A$>$1 antinuclei.

The primary purpose of the present study was to calculate the flux of antimatter particles 
of galactic and atmospheric origin, anchoring them as firmly and as widely as possible in the 
existing body of experimental data.
The galactic and atmospheric production of \pb and \db, as well as \tb, \hetb, and \heqb,
and their flux in orbit near earth can be evaluated at so far unmatched level of confidence
by using the available hadronic production data for these particles on the one hand, and the 
proven coalescence model on the other hand. 
A few preliminary steps of this work have been covered already and published: 
The systematics of the \pb data over a wide range of incident momentum has been recently 
gathered and reanalyzed \cite{DU03}, providing an accurate parametrization of the inclusive 
antiproton production cross section. The first calculations of the atmospheric \pb flux were 
reported in \cite{HU03}. Other steps dealing with the coalescence models are quoted below.
In addition, some important aspects of the production mechanisms of light composite nuclei 
and antinuclei relevant to the study are available from high energy heavy ion physics data, 
strengthening its grounds~: a key point is that the deuteron coalescence parameter derived 
from empirical fits to the data has been found to be energy independent to within a few tens 
of percents, through a wide incident energy range including the range of interest here, for 
$pA$ systems (see \cite{AR99a,SG95} for example, and below). 
This feature is expected to be valid for antideuteron coalescence as well, on general 
theoretical grounds. It is also supported by experimental indications \cite{BE99,VB99,AM98}.

The rescattering on the ISM (or on the atmospheric nuclei) of the secondary particles in the 
propagation, and the contribution of this process to the low momentum flux of CR particles is 
an important aspect of the calculations since several fluxes of primary origin and of major 
astrophysical interest are expected to contribute to the low momentum range, where the knowledge 
of the secondary flux has then to be most accurate. The issue has been addressed in this work, 
with the same concern of keeping as close as possible to the experimental facts as well as 
following general principles. It is discussed in a dedicated section and appendix below. 
The subject of \db rescattering has been addressed previously in ref~\cite{DBGLX}.

Finally, whereas the atmospheric antinuclei fluxes depend mostly on the proton and helium 
fluxes at the top of atmosphere (TOA), which are well measured, the galactic secondary 
antinuclei fluxes must be evaluated by means of a propagation model. Two classes of such models 
are widely used. The simplest one is the popular leaky box model (LBM), whereas a more realistic 
approach is the two-zone diffusion model (DM) \cite{DO01}. In the former, all the details about the 
transport conditions are discarded. These are globally accounted for by one single parameter, the 
confinement time of Cosmic Rays in the Galaxy, which, once fixed, leads to a balance between the 
production of secondary species (LiBeB, Z=21-23) from their respective primary parents (CNO, Fe) 
and the escape from the Galaxy, to reproduce the data \cite{MA01}. This model gives correct 
results for the particle fluxes of stable nuclear species.

The paper is organized as follows. The hadronic production cross sections and the coalescence 
model are discussed in section~\ref{XSEC}. The total reaction cross sections used in the 
calculations for the antimatter absorption are presented in \ref{SIGRABA}, with the particles 
rescattering issue treated in section \ref{NAR}. The antimatter propagation is discussed in 
section~\ref{GALAC}, while the atmospheric production is presented in the following 
section~\ref{ATMO}. The results are discussed and the work is briefly concluded in the final 
section~\ref{CONC}. Some details on the physics of the rescattering process are given and 
discussed in the appendix.


\section{Antimatter nuclei production cross sections}\label{XSEC}
  \subsection{Production mechanism and coalescence process}\label{COALM}
  \subsubsection{The standard coalescence model}
In the standard coalescence model, the invariant differential cross section for the (inclusive)
production of composite fragments with nuclear mass number $A$, is related to the inclusive
production cross section of nucleons by a simple power law \cite{BU63}:
%
\begin{equation}
(E_A\frac{d^3N_A}{d\vec{p}_A^{\,3}})=B_A\cdot(E_p\frac{d^3N_p}{d\vec{p}_p^{\,3}})^A ,
\label{COAL0}
\end{equation}
%
where $N_A$ and $N_p$ are the A nucleus and nucleon production multiplicities respectively,
and with $\vec{p}_A=A\cdot\vec{p_p}$. The coalescence coefficient $B_A$ can be expressed in
terms of the coalescence momentum $p_0$ and of the nuclear characteristics of the colliding
system.
%
%

In the case of deuteron or antideuteron coalescence out of matter-antimatter production, $B_A$ is
defined by the relation giving the probability of finding a neutron and a proton within $p_0$ distance
in the momentum space \cite{BU63,CS86}, which leads to:
\begin{equation}
\gamma\frac{d^3N_d}{d\vec{p}_d^{\,3}}=\frac{4\pi}{3}p_0^3
(\gamma\frac{d^3N_p}{d\vec{p}_p^{\,3}})
(\gamma\frac{d^3N_n}{d\vec{p}_n^{\,3}})
\label{COAL1}
\end{equation}
where $\gamma$ is the Lorentz factor. The multiplicity for particle $i$ is
$$\frac{d^3N_i}{d\vec{p}_i^{\,3}}=\frac{1}{\sigma_R}\frac{d^3\sigma_i}{d\vec{p}_i^{\,3}}$$
where $\sigma_R$ is the total nucleon-nucleus reaction cross section.
Assuming equal neutron $n$ and proton $p$ production cross sections and momentum distributions, the
relation can be approximated as:
\begin{equation}
\gamma\frac{d^3N_d}{d\vec{p}_d^{\,3}}=\frac{4\pi}{3}p_0^3
(\gamma\frac{d^3N_p}{d\vec{p}_p^{\,3}})^2
\label{COAL2}
\end{equation}
This relation can then be straigtforwardly extended to the coalescence of $A$ nucleons as
\cite{CS86}:
\begin{equation}
\gamma\frac{d^3N_A}{d\vec{p}_A^{\,3}}=(\frac{4\pi}{3} p_0^3)^{(A-1)}
(\gamma\frac{d^3N_p}{d\vec{p}_p^{\,3}})^A
\label{COALA}
\end{equation}
Finally, the coalescence coefficient deduced from relations \ref{COAL0} and \ref{COALA} above,
is given by:
\begin{equation}
B_A=(\frac{4\pi}{3}p_0^3)^{(A-1)}\frac{m_A}{m_p^A}
\label{BA}
\end{equation}
%

The inputs for the calculation of the production cross section for mass $A$
nuclei (antinuclei) then consist only of the value of the $B_A$ ($B_{\bar{A}}$)
parameter, the proton (antiproton) production differential cross section,
and the total $pp$ reaction cross section $\sigma_R$.


As quoted in the introduction, the deuteron coalescence momentum or coalescence radius (see
refs~\cite{CS86} for example for the definitions), derived from empirical fits to the data has 
been found to be practically energy independent between a few hundred MeV per nucleon and 158~GeV 
per nucleon incident kinetic  energy (see e.g., \cite{AR99a,SG95} and \cite{BA94,SA94,NA81,AB87,
BU80,CR75} for more detailed studies), as expected from the simple coalescence model \cite{BU63}. 
The property seems to hold as well for mass 3  nuclei and for $^4$He \cite{SA94}.
The B$_{A=2}$ and B$_{A=3}$ coalescence parameters for $p$ induced collisions on nuclei can thus 
be taken constant over this range.

In the simple coalescence model successfully describing the light nuclei production in $pA$ and 
$AA$ collisions at low and intermediate energies, the coalescence nucleons originate from the 
excited nuclear matter of the target (projectile) nucleus, and coalescence particles are produced 
in the target (projectile) rapidity range. 
In the case of high energy $pp$ collisions, nucleon-antinucleon pairs originating from the 
hadronization process are produced in the nucleon-nucleon (NN) center of mass (cm) rapidity region. 
The coalescence nuclei originate from these hadronization products which production cross 
section is maximum at rest in the cm and which therefore fly in the laboratory frame with the cm 
velocity. In these collisions, the rapidity distributions of the produced particles is thus 
symmetric with respect to the cm rapidity \cite{AR99a,AR99b,BE00,AM98,BE95}.
In high energy $AA$ collisions, both mechanisms are allowed, the rapidity distributions of light 
nuclei produced by nuclear coalescence of target nucleons are centered around the target rapidity, 
i.e., in the low momentum region in the laboratory frame for fixed target experiments, and 
symmetrically around the projectile rapidity (see \cite{AM98} for example). In these collisions, 
the coalescence production from single $NN$ collisions is also expected to exist, with a much 
lower cross section however.

There are therefore two sources of coalescence nucleons, produced in very different dynamical 
regimes: one is the production out of excited nuclear matter, the other is the production out of
hadronization products from individual $NN$ collisions. The coalescence model which applies well 
to the first case doesn't necessarily apply to the second.
In this latter reaction however, the measured $d$ and \db production cross sections have been 
shown to be both consistent with a coalescence production mechanism as for nuclear systems 
\cite{DO65,AL73b,AL75,GI78,AB84}, which indicates that the same approach can be used both for 
matter and antimatter production calculation.

In addition, the similarity of the production mechanism for light nuclei and antinuclei was also 
suggested by various experimental indications in $AA$ collisions. In this case however, the 
coalescence parameter was found to be incident energy dependent \cite{AR99a,SG95,VB99,BE99,HA02}.
Note that the microscopic approach to the coalescence model was developed initially \cite{KO92}
to explain this dependence of the coalescence parameter on the kinematic conditions of the reaction.
For $p$+$A$ systems however, the coalescence coefficients were found to be energy independent as 
mentioned above, for $d$ as well as for \db \cite{AR99a,SG95,BE00}. This will be confirmed below 
for \db particles.

These observed dynamical and kinematical features allow therefore to make reliable predictions for 
the production in space of light antimatter nuclei from $AA$ collisions ($A$ standing here for any 
nucleus including proton), over a range of incident CR proton energies conveniently matching the 
useful range for the present study.
    \subsubsection{Microscopic coalescence model}
A microscopic approach to the coalescence model has been developed recently in \cite{KO92,DU03b}, 
which successfully described the mass 2 and 3 antinuclei production 
$pA\rightarrow$\db$X$ and $pA\rightarrow$\3b$X$ (\3b standing for \tb and \hetb) 
experimental data. The model has been used in the present work as a complementary tool to 
the standard coalescence model. 

The two models have been used in the study, in order to provide an estimate of the 
theoretical uncertainty to be assigned to the calculations. This uncertainty was turned 
around however by the standard model being fitted to the available data.
  \subsection{Antiproton production cross section}\label{PBPROD}
The cross section data for the proton induced inclusive \pb production on nuclei and on the 
nucleon, have been reanalyzed recently, in the framework of a research program aiming at the 
understanding of the secondary particle flux produced in the atmosphere \cite{HU03}. This 
was achieved using an improved analytical form used in previous works for fitting the cross 
sections \cite{DU03}, and it was shown that the data could be described with a fair accuracy 
over the incident energy range between 12 and 400~GeV. The results of this work, i.e., for 
$pA\rightarrow\bar{p}X$ and $pp\rightarrow\bar{p}X$ cross sections, have been used in the 
present calculations, and its relevant part for the present study is repeated here 
for convenience. 

The parametrization of the inclusive $pA\rightarrow\bar{p}X$ 
cross section data is one of the two major ingredients of the present calculations, on 
which results and conclusions are built. The $pA\rightarrow\bar{p}X$ cross sections
obtained from the analysis of \cite{DU03} are estimated to be accurate to within about 20\% 
above $T_{\bar{p}}\approx$1~GeV. 
The low kinetic energy \pb data 0.5~GeV$\lesssim T_{\bar{p}}\lesssim $1~GeV 
are probably less accurate. They tend to be somewhat underestimated by the analytical form 
used \cite{DU03}. No fully reliable estimate of the low energy $T_{\bar{p}}\lesssim$0.4-
0.5~GeV accuracy can be assessed because of the total lack of data over this range and of the
high sensitivity to nuclear medium effects. Symmetry considerations also constrain the low 
energy data (see the discussion in \cite{DU03}) for the $pp$ system. Unfortunately, these 
cannot be fully applied to the $pA$ data because of the nuclear medium effects on the final 
state in the \pb production, which may more than significantly distort the spectrum. This 
issue would deserve a dedicated study by itself.  
For the $pp\rightarrow \bar{p}X$ cross section, two versions of the parametrization are 
available: the A=1 version of the $pA$ parametrization (below referred to as I) , and a 
dedicated simplified version of the formula with no nuclear mass dependence in the analytical 
form, which provided a markedly better fit to the \pb production in $pp$ collisions data (below 
referred to as II). The two versions provide values of the integrated cross sections (multiplicity) 
differing by about 20\%, the $[pA\rightarrow \bar{p}X]_{A=1}$ fit (I) providing the larger 
values. Both parametrizations have been used in the present work for the $pp$ cross sections.
  \subsection{Antideuteron production}\label{DBARCOAL}
Galactic antideuterons are expected to be produced in $pA$ and $AA$ collisions dominantly by 
the elementary $pN\rightarrow\bar{d}X$ process via a coalescence mechanism out of the hadronic 
production of $N\!\bar{N}$ pairs as discussed previously. Another competing production is possible 
however in space, from the $\bar{p}p\rightarrow \bar{d}X$ reaction. This reaction has to be 
considered since although the $\bar{p}$ flux is much lower than the $p$ flux, only one 
$N\!\bar{N}$ pair has to be produced in the kinematical domain where the "incident" $\bar{p}$ 
in the final state can coalesce with a $\bar{n}$ particle produced in the collision, providing
a lower energy \db flux than the other reaction but with a comparable magnitude. 
Both production channels are discussed in the next two subsections.

  \subsubsection{Antideuteron production in the $pA\rightarrow \bar{d}X$ reaction}\label{PPDB}

\begin{figure}[htbp]          
\begin{center}
\includegraphics[width=\columnwidth,angle=0]{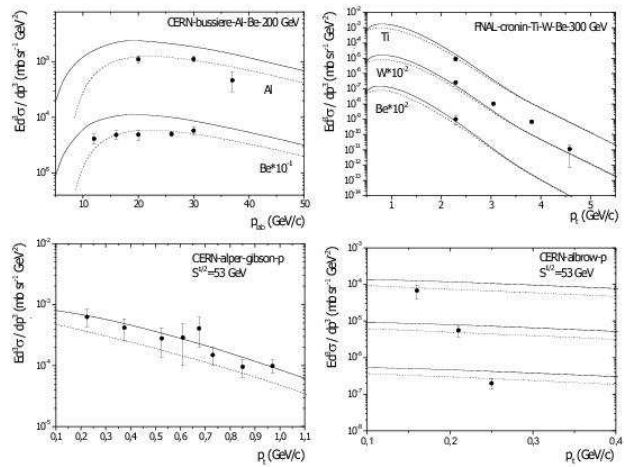}
\caption{\it\small Comparison of the deuteron production cross section calculated with the standard 
(dashed lines) and microscopic (solid lines) coalescence models, with the experimental data 
from pA collisions at lab momenta 200~GeV/c on Be and Al targets (top left) \cite{BU80}, 
at 300 GeV/c on Be, Ti and W  targets (Top right) \cite{CR75}, and at 53~GeV center of mass 
energy on the proton from \cite{GI78} (full circles) and \cite{AL73b} (open circles, bottom 
left), and from \cite{AL75} (bottom right).
\label{COALD}}
\end{center}
\end{figure}

The simple coalescence model described above has been applied to a set of antideuteron cross 
section production data available in the literature \cite{BU80,CR75,GI78,AL73b}. 
Fig.~\ref{COALD} shows the results obtained using the coalescence formula \ref{COALA} with
the \pb spectra obtained as in \cite{DU03}, and fitting the value of the coalescence momentum 
to a selected sample of data, with the result $p_0$=79~MeV/c. 
As seen on the figure the data are well reproduced by the calculations over a range extending
from 200~GeV/c incident momentum up to the ISR center of mass (cm) energy s$^{1/2}$=53~GeV.
In total 34 data points have been fitted. The results are given in table~\ref{FITDB}.

\begin{figure}[htbp]          
\begin{center}
\includegraphics[width=\columnwidth,angle=0]{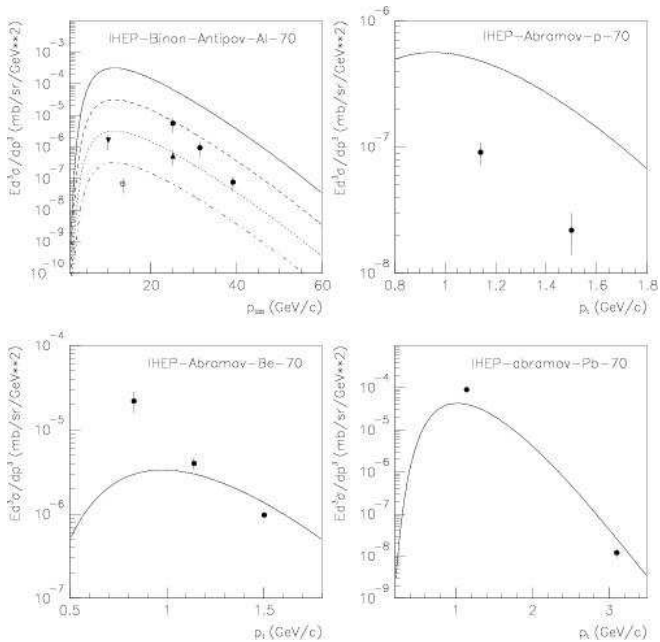}
\caption{\it\small Comparison of the antideuteron production cross section calculated with 
the standard coalescence model as described in the text, with the inclusive experimental data 
from pA collisions at 70~GeV/c from refs \cite{AN71,AB84,AB87}. The upper left frame shows 
the data measured at a set of angles, 0$^{\circ}$ (full circle, solid line), 12~mrad (inverted 
triangle, dashed line), 27~mrad (full triangle, dotted line), and 47~mrad (open circle, 
dash-dotted line). See text for details.
\label{BADDB} }
\end{center}
\end{figure}

The 70~GeV data from refs \cite{AN71,AB87} have been discarded from the search. They 
were found not to be compatible with the other data, since they provided exceedingly large 
$\chi^2$ values. These data are listed in table~\ref{NOFITDB}. It is seen in the table that 
the data measured on $Pb$ target have a not so bad $\chi^2$ per point value (bottom line in 
the table). They have not been used in the analysis however, for the sake of consistency.
Although the observed disagreement could be of physical origin since the discarded data have 
been measured at an incident energy below the momentum range of good agreement in 
Fig.~\ref{COALD}, they have not been treated on the same footing as the other data because 
the \pb yields measured in the same set of experiments were already found to be inconsistent 
with the measurements from other works \cite{DU03}. The comparison of these data with the 
standard coalescence model calculations using a coalescence parameter fitting the data sets 
listed in table~\ref{FITDB}, are shown on Fig.~\ref{BADDB}. The disagreement clearly varies 
from moderate to sharp, with no clear trend pointing to some possible physical origin. The 
data are very overestimated or very underestimated by the calculations for the same incident 
energy for the different measurements. 
%
\begin{table}
\begin{center}
\begin{tabular}{lcccc}\hline\hline
Experiment & target & $p_{inc}$ or $\sqrt{s}$& $N$  & $\chi^{2}$/N \\
           &        &  GeV/$c$      GeV   &                   &                      \\
\hline
Bussiere \textit{et al}.
\cite{BU80} & $Be$  & 200 & 5 & 0.95  \\
            & $Al$  & 200 & 3 & 1.7  \\
Cronin \textit{et al}.
\cite{CR75} & $Be$  & 300 & 1 & 4.4  \\
            & $Ti$  & 300 & 1 & 1.3  \\
        & $W$   & 300 & 4 & 13.7  \\
Alper \textit{et al}.; Gibson \textit{et al}. \cite{GI78}  & $p$  & $\sqrt{s}=53$ & 8 & 1.3  \\
Albrow \textit{et al}.
\cite{AL75}  & $p$ &$\sqrt{s}=53$ & 3 & 1.0  \\
Armitage \textit{et al}.
\cite{AR79}  & $p$ &$\sqrt{s}=53$ & 9 & 4.6  \\
\hline \hline
\end{tabular}
\caption{\small
Experimental \db production cross section data used in the coalescence model analysis to fit 
the coalescence momentum parameter as discussed in the text. 
$N$ is the number of experimental points used in the fit. \label{FITDB}}
\end{center}
\end{table}
%
\begin{table}
\begin{center}
\begin{tabular}{ccccc}\hline\hline
Experiment & target & $p_{inc}$ & $N$  & $\chi^{2}$/N \\
           &        &(GeV/$c$)  &      &                      \\
\hline
Binon \textit{et al}.;
Antipov \textit{et al}. \cite{AN71} & $Al$ &70 & 6 & $2.10^{4}$  \\
Abramov \textit{et al}. \cite{AB84,AB87} & $p$ &70 & 2 & $1.10^{3}$ \\
                & $Be$ &70& 3 & $3.10^{2}$  \\
                & $Pb$ &70& 4 & 15.4  \\
\hline \hline
\end{tabular}
\caption{\small Experiments whose results were not included in the search. The chi-squared 
per point values (rightmost column) were obtained using the coalescence parameters fitted to 
the selected data listed in table \ref{FITDB}. Same definitions as in previous table.
 \label{NOFITDB}}
\end{center}
\end{table}
In the calculations of the \db and \3b yields, the threshold effects were taken into account 
by means of the phase space of the final state as described in ref~\cite{DU03b}.

  \subsubsection{Antideuteron production in the $\bar{p}A\rightarrow \bar{d}X$ reaction}\label{PPBDB}

\begin{figure}[htbp]          
\begin{center}
\includegraphics[width=\columnwidth,angle=0]{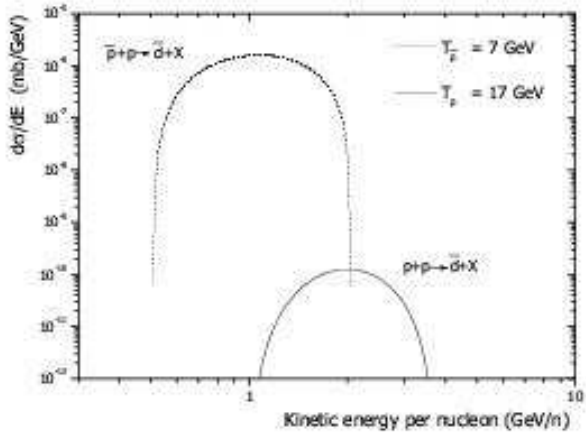}
\caption{\it\small Spectral distribution of the antideuteron production cross section in the two 
reaction schemes $pp\rightarrow\bar{d}X$ (solid line) and $\bar{p}p\rightarrow\bar{d}X$ (dashed 
line) considered, calculated with the standard coalescence model as described in text. $T_{\bar{p}}$
and $T_p$ are the production thresholds. 
\label{COALSD} }
\end{center}
\end{figure}

For the antideuteron production calculation in the $\bar{p}p(He)\rightarrow\bar{d}X$ 
reaction(s) either in the simple or in the microscopic, coalescence model, the inclusive \pb 
production cross section from the same incoming channel is needed. This cross section 
being unknown experimentally, an evaluation procedure has been used to estimate it on the 
basis of reliable approximations. In the assumed production mechanism the 
$\bar{p}p\rightarrow\bar{d}X$ reaction produces one single $N$\Nb pair out of which the \nb 
(antineutron) will coalesce with the 'existing' \pb in its final state, into a \db$\!$, 
i.e., $\bar{p}p\rightarrow(\bar{p}\bar{n})npX\rightarrow\bar{d}X$. 
The following assumptions based on the knowledge of the $\bar{p}p\rightarrow\bar{p}X$ and
$\bar{p}p\rightarrow\bar{n}X$ processes, have been made for this evaluation:
\begin{enumerate}
\item The probability to produce an antiparticle by hadronisation in a \pb$\!p$ collision 
$\bar{p}p\rightarrow\bar{n}X$ in the final state energy range of interest, i.e., excluding 
the quasi elastic charge exchange processes to the resonance region, is the same as the 
probability to produce a \pb in a $pp$ collision $pp\rightarrow\bar{p}X$:
%
\begin{equation*}
E_{\bar{n}}\frac{d^3\sigma_{\bar{n}}}{d^3p_{\bar{n}}}(\bar{p}p\rightarrow\bar{n}X)
\approx
E_{\bar{p}}\frac{d^3\sigma_{\bar{p}}}{d^3p_{\bar{p}}}(pp\rightarrow\bar{p}X).
\end{equation*}
For the $\bar{p}p(H\!e)\rightarrow\bar{n}X$ inclusive cross section, parameterisations
II and I were used for $pp$ and $pHe$ collisions respectively. 
\item The $\bar{p}p\rightarrow\bar{p}X$ cross section is assumed to be the same as the 
$pp\rightarrow pX$ cross section, namely:
\begin{equation*}
E_{\bar{p}}\frac{d^3\sigma_{\bar{p}}}{d^3p_{\bar{p}}}(\bar{p}p\rightarrow\bar{p}X)
\approx
E_{\bar{p}}\frac{d^3\sigma_{\bar{p}}}{d^3p_{\bar{p}}}(pp\rightarrow pX)
\end{equation*}
This approximation makes sense since the collision dynamics are similar for the two systems 
at high enough collision energies. It is probably not correct however for low energy 
secondary \pbs from nuclear targets because of the dominance of the annihilation cross 
section in \pb$N$ collisions at these energies and of the subsequent interactions of the
produced \pbs with the nuclear medium. 
For the proton inclusive cross section, the parameterisation from \cite{KMN} has been used.  

An additional correction has to be made for the energy dependence of the \db production 
cross section near the threshold, assumed to follow the phase space available to the 
particle \cite{DU03b}. 
\end{enumerate}
The differential cross sections for the $pp\rightarrow\bar{d}X$ and $\bar{p}p\rightarrow
\bar{d}X$ reactions calculated using the coalescence model(s), are compared on 
Fig.~\ref{COALSD} for incident energies near their respective threshold. For the latter 
reaction, the calculation was performed within the approximations described above.
It can be seen on the figure that the $\bar{p}p\rightarrow\bar{d}X$ cross section is much 
larger than for the $pp\rightarrow\bar{d}X$ reaction, the maximum values being about four 
orders of magnitude apart. The distributions are also centered at very different particle 
energies, corresponding to the respective NN center of mass velocities in both cases, and 
being thus much lower for the $\bar{p}p\rightarrow\bar{p}p(\bar{n}n)$ reaction than for 
$pp\rightarrow pp(\bar{p}p)(\bar{n}n)$ reaction.
These features clearly show that the $\bar{p}p\rightarrow\bar{d}X$ contribution has to be 
taken into account in the calculations of the galactic \db flux.

  \subsection{Mass 3 and 4 antinuclei production cross section}\label{ABPROD}

The production cross sections of mass 3 antinuclei (\tb,\hetb) in $pA$ collisions are 
extremely scarce. They have been measured in the past only at 70~ GeV \cite{AN71b,VI74}, and 
at 200 and 240~GeV \cite{BU80}.

In the Serpukhov experiments \cite{AN71b,VI74}, the \tb and \hetb production cross sections 
have been measured at small production angles on $Be$ and $Al$ targets. The \pb and \db data 
from the same series of experiments have been discarded from the sets of data fitted in the 
present analysis, as discussed previously. For the results of the mass 3 antinuclei production 
cross section measured in these experiments, no consistent interpretation could be found either 
\cite{DU04}, albeit the standard coalescence calculations are found in reasonable agreement with 
the data, i.e., typically less than one order of magnitude larger.

\begin{figure}[htbp]          
\begin{center}
\includegraphics[width=\columnwidth,angle=0]{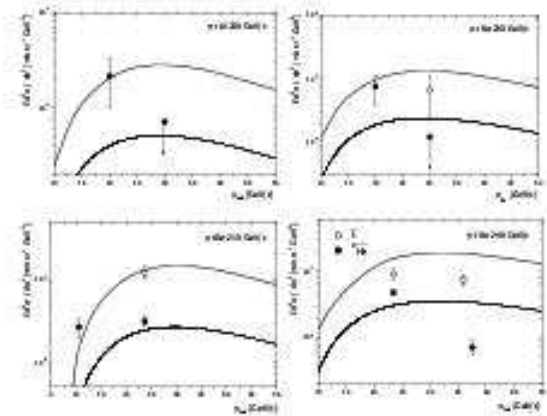}
\caption{\it\small Comparison of the production cross section for mass 3 antinuclei calculated 
by means of the standard (dashed line) and microscopic (solid line) coalescence model as 
described in the text.
\label{3BAR} }
\end{center}
\end{figure}

The production cross sections of mass 3 antinuclei have also been measured at CERN in a series 
of two experiments \cite{BU80}. The measured \pb and \db cross sections have been discussed in the 
previous subsection. The mass 3 data are compared on Fig.~\ref{3BAR} to the calculated cross 
sections in the coalescence models.  
The calculated cross sections are found within one order of magnitude from the data points for the 
two studied targets, with the two coalescence models providing values approximately framing the 
data points, the microscopic model providing the upper limit. 
This agreement can be considered as fair as far as the order of magnitude is concerned. It is quite 
a significant result for the present study which purpose is to fix the orders of magnitude of the 
corresponding secondary CR flux.
It is interesting to note on the figure, that the experimental production cross section is found 
significantly larger for \tb than for \hetb for all values of momenta where both have been measured. 
Should this difference be assigned to Coulomb effects in the coalescence process is an open question 
which would deserve a dedicated investigation. It must be noted however than the production cross 
sections for $t$ and \het particles measured in the same experiments were found almost identical 
\cite{BU80}. 
\section{Other cross sections required for the calculations}\label{DBPROD}
  \subsection{Total antimatter reaction cross sections on nuclei}\label{SIGRABA}
The total reaction cross sections $\sigma_R(\bar{a}A)$ for light antinuclei \ab collisions 
on nuclei $A$ are needed for computation of the absorption term in the LBM or DM transport 
equations.
    \subsubsection{Antiprotons}\label{CSRPB}
\begin{figure}[htb]          
\begin{center}
\includegraphics[angle=0,width=\columnwidth]{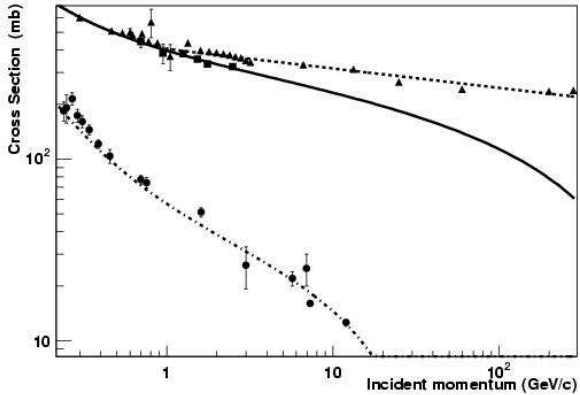}
\caption{\it\small Total \pb reaction cross section on the on $^{12}C$ (full triangles, see 
\cite{BE94} for the sources of data), compared with the fit using relation \ref{SIGRPBA} 
(dashed line). The \pb$\!C$ annihilation data (full squares) are compared with a fit using 
relation~\ref{SIGAPBA} (solid line).
The \pb$\!p$ annihilation data (full circles) are compared with the fit to the data points 
from \cite{TA83} (dot-dashed curve). 
\label{FCSPBARA}}
\end{center}
\end{figure}
For the galactic propagation, the $\sigma_R(\bar{p}p)$ total reaction cross sections for 
the \pb$\!\!p$ collisions were calculated following \cite{TA83}, while for \pb$^4\!He$ 
collisions they were calculated from the former in the Glauber approximation as described 
in the next section (see \cite{DU04} for more details).

For the atmospheric propagation of \pbs in the atmosphere, the total reaction $\sigma_R$ and 
annihilation $\sigma_{ann}$ cross sections for $\bar{p}\!A$ collisions on light nuclei ($^{14}N$ and 
$^{16}O$) were calculated using the following formulas (in mb):
%
\begin{eqnarray}
\sigma_R=(257.8+\frac{88.7}{T})(\frac{A}{12})^{2/3}         \label{SIGRPBA} \\
\sigma_{ann}=0.661(1+0.0036T^{-0.774}-0.902T^{0.0151})A^{2/3}    \label{SIGAPBA}
\end{eqnarray}
%
$T$ being the particle kinetic energy. This relation is based on a similar formula given in 
\cite{BE94} (p 39 of this ref.), with the $A^{2/3}$ dependence added. The resulting fit is 
shown on Fig.~\ref{FCSPBARA}.
    \subsubsection{Antideuterons}\label{CSRDB}
 A few experimental data for the total reaction cross section $\sigma_R($\db$A)$ are 
available \cite{BI70,GO72}. The same functional dependence on the nuclear mass number $A$ as 
obtained in \cite{BI70} has been used in this work:
\begin{equation}
\sigma_R=105 A^{2/3}\,\,mb
\label{SIGRDB}
\end{equation}
\begin{figure}[htb]          
\begin{center}
\includegraphics[width=\columnwidth,angle=0]{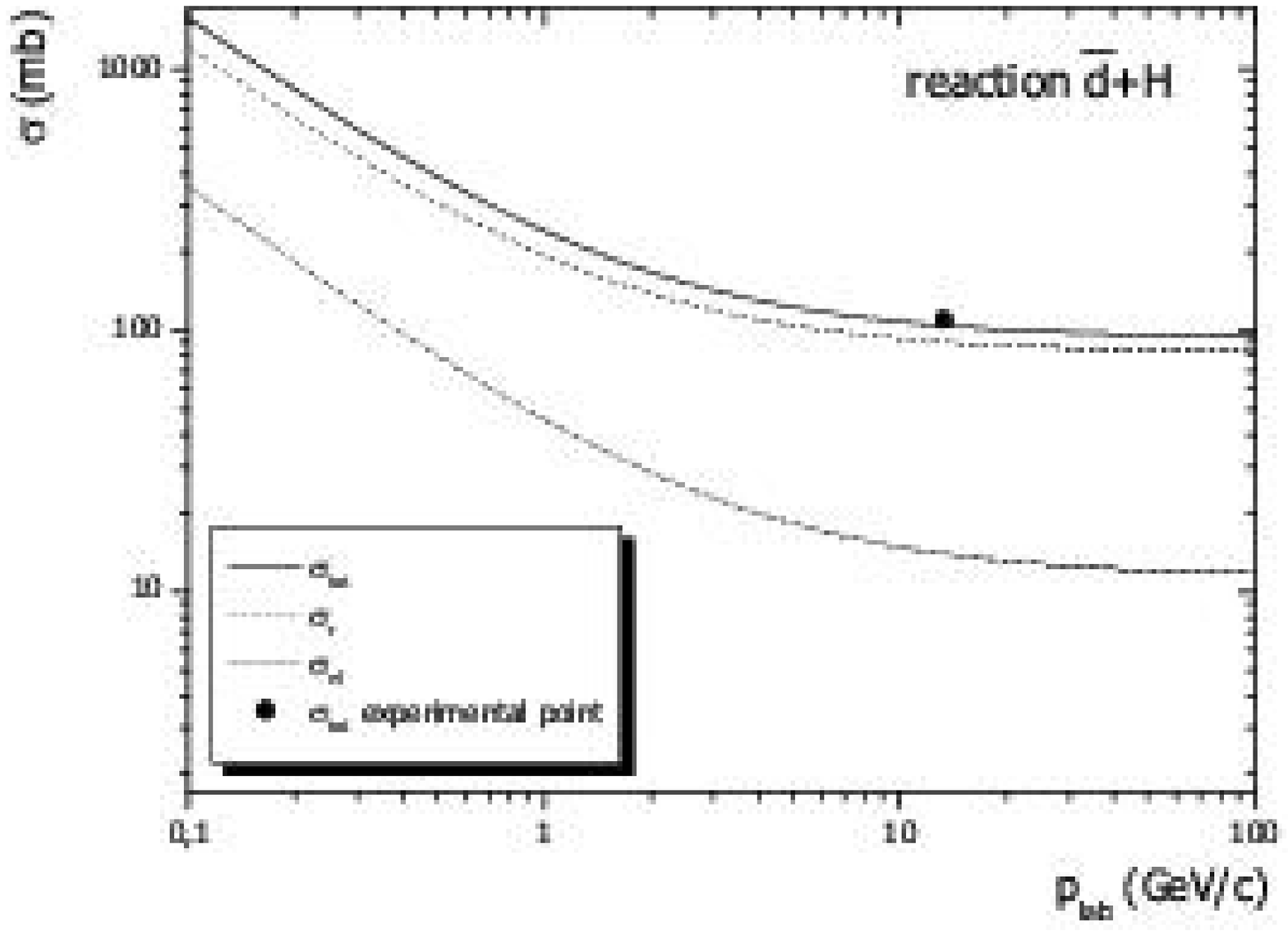}
\includegraphics[width=\columnwidth,angle=0]{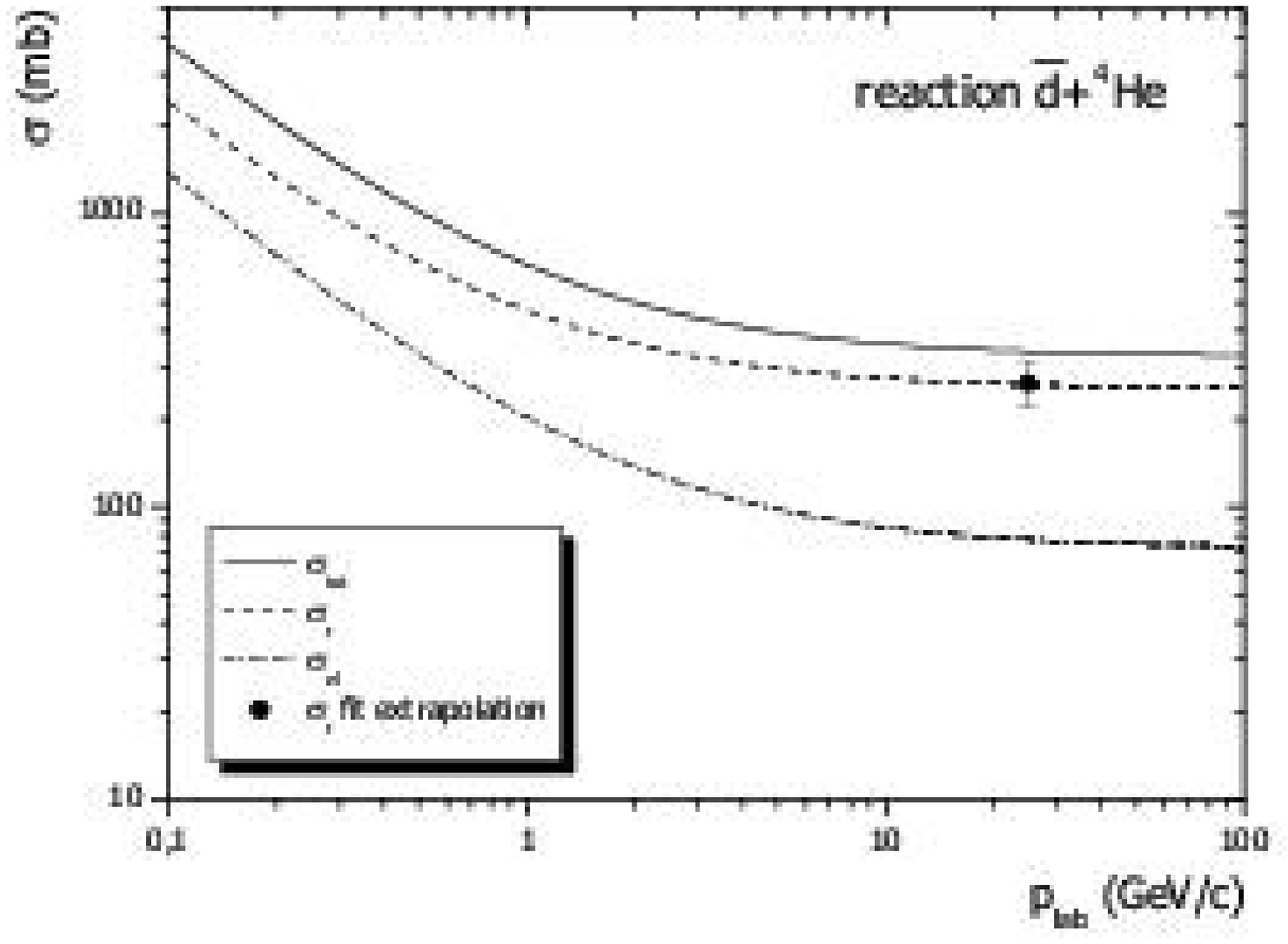}
\caption{\it\small
Top: Total \db$\!p$ cross section data from \cite{GO72} compared with the calculated values 
in the single scattering approximation of the Glauber model, as explained in the text. Note 
the good agreement between calculations and data.
Bottom: Comparison of the total \db$^{\!4}\!He$ reaction cross section obtained using the 
empirical relation~\ref{SIGRDB} (full circle, the error bar is from \cite{BI70}), with the 
Glauber approximation results (dashed line). The calculated total (full line) and elastic 
scattering (dash-dotted line) cross sections are also given, for completeness. See text for 
details.
\label{SIGDBPHE}}
\end{center}
\end{figure}
The total reaction cross sections have also been calculated using the Glauber approximation 
and the parameters of the elementary \pb$\!p$ scattering amplitude using the $\bar{N}N$ 
amplitudes from the analysis of \cite{FU90}.
The results are compared on Fig.~\ref{SIGDBPHE} with the experimental results on the proton
target and with the empirical parametrization from relation \ref{SIGRDB}. See \cite{DU04} for
more details.
    \subsubsection{Mass 3 and 4 antinuclei}
For the mass 3 and 4 antinuclei, the total reaction cross section has been calculated in the 
same eikonal approach as for the \pb$\!A$ and \db$\!A$ systems, using the known matter 
distribution of the colliding nuclei and the nucleon-antinucleon amplitudes \cite{DU04,
FU90}.

\subsection{Non annihilating inelastic rescattering} \label{NAR}
Before proceeding to the calculations of the propagated fluxes, the important issue of the
rescattering of the transported particles is discussed in this section and more extensively
in the appendix.

The low kinetic energy range - $T\lesssim$500~MeV per nucleon - antideuteron flux is most sensitive
to the possible primary \db flux originating from exotic astrophysical sources such as
primordial black holes \cite{DBPBH} or dark matter (neutralinos) annihilation \cite{DBDM}.
Any calculations of the secondary flux aiming at a good accuracy for low energy particles 
must then take into account all the significant effects contributing to populate this 
energy-momentum region. Note that this argument holds qualitatively as well for antiprotons 
or any other particle or antiparticle propagating in the interstellar medium. 
The rescattering of particles involves some energy loss and thus the transfer of a 
fraction of the flux from a given energy to a lower energy. 
The energy loss induced by the elastic scattering of the transported particles at the
energies considered here involves only small momentum transfers. It is negligible.
The inelastic process \db$\!p\rightarrow$\db$\!X$ may involve large energy-momentum loss of
the scattered particles.
A reliable description of the secondary flux in the low energy region should thus take into
account the component induced by the \db$\!p\rightarrow$\db$\!X$ Non Annihilating inelastic
(Re)scattering (NAR) reaction of the particles propagated in the ISM.
This rescattering component will be referred to as tertiary in the following.
  \subsubsection{Principles}
It has been argued recently that \db (or symmetrically d) particles incident on a nucleon or
on a nuclear target should have a small inelastic scattering cross section because of their
natural "fragility" originating in the small \db(d) nuclear binding energy, since a deuteron 
bound by only 2.2~MeV easily dissociates in a collision.
This intuitive argument is misleading however, and appears not to be correct after a careful 
examination of the problem. It can be invalidated both on empirical evidence and on formal 
grounds. The former consists of the existing experimental evidence of deuteron \cite{BA73,
GO79} - and more generally nuclei \cite{MO92} - induced nucleon excitation cross section 
(see also section \ref{NARpractical} below). 
The latter requires a few developments given in the appendix where it is shown on general 
grounds that the inelastic scattering cross section of light nuclear systems is in fact 
expected larger than the corresponding ($p,p'$) or ($\bar{p},\bar{p}'$) cross section, 
basically on the simple argument that it is driven by the $NN$(\Nb$\!$\Nb) elementary cross 
section folded with the matter distribution of the colliding systems. This is in agreement 
with the available experimental facts.

{\bf Effects of the isospin selection rules} \\
This picture is somewhat blurred however in the case of the inelastic (d,d') reaction on the
nucleon since the inelasticity is strongly inhibited at low incident energies by the isospin
selection rule, the  isovector excitations in the target nucleon being forbidden for this
reaction.
The deuteron is an isoscalar particle with isospin 0 which cannot induce isovector (i.e.,with 
isospin quantum number T=1) excitations in a nucleon ($\Delta$ resonances), in single step 
transitions.
This has a strong inhibition effect on the total inelasticity of the reaction since the first
excited state of the nucleon is the (isovector) $P_{33}\;\Delta$ resonance which in addition
largely dominates the excited nucleon spectrum when the transition is permitted in inelastic
processes like (e,e') or (p,p'). No direct nucleon excitation is thus permitted in (d,d')
below the first (isoscalar) N$^*$ resonance in hadron collisions, namely below the 1.4~GeV 
(Roper) and 1.52~GeV \cite{AL73a} resonances in the nucleon.

The overall inelasticity is not totally hindered however since two step excitations
via the nucleon ($\Delta$) resonance(s) in the deuteron are allowed and have been observed
experimentally \cite{BA73,CE76}.
  \subsubsection{Practical method}\label{NARpractical}
\begin{figure}[htb]          
\begin{center}
\includegraphics[width=\columnwidth,angle=0]{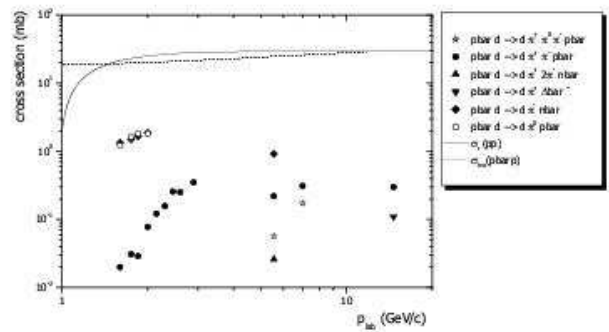}
\caption{\it\small Partial \pb$d\rightarrow (n\pi)$\pb$d$ cross sections from \cite{BA88}
used to evaluate the total inelastic NAR \db$p\rightarrow$\db$X$ cross section.
\label{SIGDBP}}
\end{center}
\end{figure}

In account of the complex interplay between the underlying NN cross sections and selection
rules which govern the \db$\!p$ inelastic cross section as discussed above, the practical 
approach used here has been based on empirical grounds. The energy integrated 
$\bar{d}p\rightarrow\bar{d}X$ cross section has been inferred from the experimental values 
of the cross section for symmetric system \pb$d\rightarrow Xd$ which should be identical at 
the same center of mass energy \cite{GO72}.
Fig.~\ref{SIGDBP} shows the experimental values for \pb$\!d\rightarrow (n\pi)$\pb$d$ ($n$ 
multiplicity of produced pions) reactions as a function of the incident energy \cite{BA88}.
The total inelastic NAR cross section $\sigma_{in}($\pb$\!d)$ has been obtained by summing 
up the \pb$d\rightarrow (n\pi)$\pb$d$ cross sections experimentally available, leading 
to $\sigma_{in}($\pb$\!d)\approx 4 mb$. 
No attempt was made to evaluate the (expectedly small) contributions of the channels not known 
experimentally. The overall evaluation is thus quite conservative. 
Note that these data also support the arguments in favor of a non negligible \db$(d)$ 
inelastic cross section, given in the previous section.
For the momentum spectrum of inelastically scattered particles, the same functional form as 
in \cite{AN67} was used to fit the data measured in these works, where the $pp\rightarrow 
pX$ inclusive cross section was found to be independent of the longitudinal momentum in the 
center of mass $p_l^*$, and could be described in the laboratory as:
\begin{equation}
\frac{d^{2}\sigma(pp\rightarrow pX)}{dpd\Omega}=\frac{p^{2}_{p}}{2\pi
p_t} \frac{\gamma(E-\beta p\cos\theta)}{E}610p_t^{2}
e^{-\frac{p_t}{0.166}} \label{eq:5-25}
\end{equation}
where $\gamma$ et $\beta$ are the usual Lorentz factor and particle velocity and $p_t$
the transverse momentum of the particle.
The integrated cross section was normalized to the value determined as above for a given \db
incident energy.

\begin{figure}[htb]          
\begin{center}
\includegraphics[width=\columnwidth,angle=0]{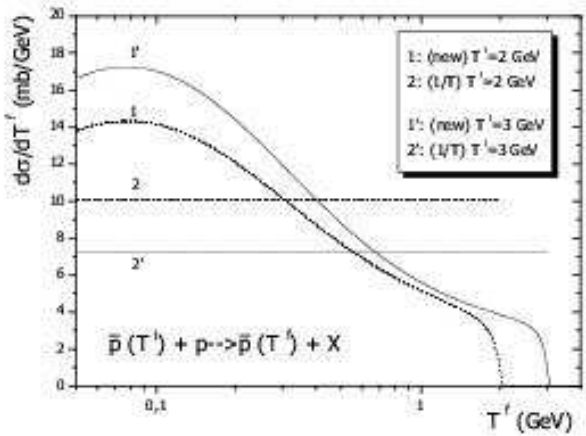}
\vspace{-0.5cm}
\caption{\it\small Spectral distributions of inelastically scattered protons (antiprotons)
from a hydrogen target in the approach of \cite{FI74} (horizontal lines) and 
used in this work for evaluating the NAR contributions of the antinuclei flux, for 2~GeV 
(dashed lines) and 3~GeV (solid lines) incident kinetic energies. 
\label{spectrNAR}
}
\end{center}
\end{figure}

The ansatz used in previous works for the energy distribution of the secondary particles
was based on the limiting fragmentation hypothesis \cite{FI74}. The form used for the energy 
dependence of the differential cross section $d\sigma(E,T_0)/dE\sim 1/T_0$ just corresponds 
to a constant differential cross section over the energy range of the produced particles 
between 0 and the incident kinetic energy $T_0$ in the laboratory frame.
Experimentally, the $pp\rightarrow pX$ differential cross section was shown to be largely 
independent of the longitudinal momentum $p_l^*$ of the produced particles in the 
center of mass \cite{AN67} (see also \cite{AL73b}). 
Fig.~\ref{spectrNAR} shows the inelastic scattering spectra in the laboratory for 2~GeV and 
3~GeV incident kinetic energy protons obtained in the approximation of \cite{FI74} (flat 
spectra) and used in this work, after \cite{AN67}. The effects of the observed differences 
on the galactic flux of particles are discussed in the section~\ref{TERTIARY} below. 
%


%
%
\section{Galactic antimatter production and propagation}\label{GALAC}

In the previous sections, the hadronic and nuclear physics issues related to light antinuclei 
production and interactions have been reviewed and updated. The present section addresses 
now the flux propagation of the produced antimatter nuclei in the framework of the leaky box 
model. The model is simple, widely successful, and highly useful, with a sound predictive 
power for use with purposes such as the present one. It rests on a single effective 
phenomenological parameter - the escape length $\lambda_{esc}$ \cite{JO01} - incorporating 
the physics (diffusion and convection) of the transport process.  
The leaky box model avoids lengthy discussions about what should be the value of the galactic 
transport parameters in more realistic models. It is an economical way to obtain reliable  
results. As such, it is very well suited to address the impact of the cross sections 
obtained here and in \cite{DU03} on the \pb and \db fluxes.

The errors on the \hetb, \heqb and to a lesser extent, \db production, are the dominant 
source of uncertainty. Even for \pbs, the inaccuracy on the production cross section has 
been estimated to be larger than that related to the propagation parameters in \cite{DO01}. 
For heavier antinuclei $\bar{A}$, the production rate is proportional to $p_0^{\bar{A}-1}$. 
An uncertainty of $\Delta p_0$ on the coalescence parameter leading to an uncertainty 
$\Delta\Phi_{\bar{d}}$ on the antideuteron flux translates into an uncertainty 
$(\bar{A}-1)\Delta\Phi_{\bar{d}}$ for these anti-nuclei.

The following subsections compare in details the \pb and \db fluxes with previous results 
for the various contributions from nuclear reactions, diffusive reacceleration, and solar 
modulation. The results derived in the simple LBM are then compared to more realistic 
calculations performed in the DM framework.
The section also includes the presentation of \hetb and \heqb fluxes and ends with a summary 
of the calculated fluxes.

\subsection{The Leaky Box Model for Cosmic Ray propagation}

In this model, the flux $\Phi_{\bar{A}}(T_{\bar{A}})$ of the antinucleus $\bar{A}$ at 
kinetic energy $T_{\bar{A}}$ is given by:
\begin{eqnarray}
\Phi_{\bar{A}}(T_{\bar{A}})=\frac{\lambda_{\rm esc}(T_{\bar{A}})\lambda_{\rm int}(T_{\bar{A}})}
{\varrho_{\rm ISM}\left[\lambda_{\rm esc}(T_{\bar{A}})+\lambda_{\rm int}(T_{\bar{A}})\right]}
\hspace{2cm}\nonumber\\
\times \frac{1}{4\pi}\left[Q_{\bar{A}}^{\rm sec}(T_{\bar{A}})+Q_{\bar{A}}^{\rm ter}(T_{\bar{A}})
\right]\label{eq:5-11}
\end{eqnarray}
with $Q_{\bar{A}}^{\rm sec}(T_{\bar{A}})$ and  $Q_{\bar{A}}^{\rm ter}(T_{\bar{A}})$ being 
the secondary and tertiary source terms discussed below, respectively, and where 
$\lambda_{\rm int}(T_{\bar{A}})=\langle m \rangle/\langle\sigma^{\bar{A}+{\rm ISM}}_R(T_
{\bar{A}}) 
\rangle$
and $\lambda_{\rm esc}(T_{\bar{A}})$ are taken from~\cite{WE96}. Following~\cite{SI98}, 
we use the quantities $\langle m \rangle =2.05\times 10^{-24}$~g for the average mass of the 
interstellar gas, $\varrho_{ISM}=2.28\times 10^{-24}$~g~cm$^{-3}$ for the ISM average density
and $\langle \sigma^{\bar{A}+{\rm ISM}}_R(T_{\bar{A}}) \rangle$ stands for the average 
reaction cross section on this gas.
The ISM composition used is ${\rm H}:{\rm He}:{\rm C}:{\rm N}:{\rm O}=1:0.1:5\times 10^{-4}:
8\times10^{-4}:8\times 10^{-5}$~cm$^{-3}$. Actually, these numbers as well as the average 
density are not perfectly known. The requirement to fit the B/C ratio for a given choice for 
these quantities, leads to a peculiar parametrization for $\lambda_{\rm esc}(T_{\bar{A}})$. 
However, a different choice than the above, if it would probably affect the normalisation of 
the fluxes, would have a minor effect on the shape of the distribution. 
It is thus sufficient to check that the choice made for $\langle m \rangle$, $\varrho_{ISM}$ 
and $\lambda_{\rm esc}$ leads to the correct normalisation of the spectra. 
Tested on the \pb spectrum, the same set of parameters can then be safely applied to the 
other antinuclei fluxes.

Concerning the secondary $Q_{\bar{A}}^{\rm sec}$ and tertiary $Q_{\bar{A}}^{\rm ter}$ source 
terms in relation~\ref{eq:5-11} above, the former results from the net creation of the 
antinucleus $\bar{A}$ from CRs interaction on the ISM (see Sec.~\ref{XSEC}), whereas the 
latter results from the energy redistribution through inelastic non-annihilating (NAR) 
reactions of the produced antinuclei, discussed in details in Sec.~\ref{NAR} above. Note that 
neither the ionisation losses nor the reacceleration process were included in the LBM 
calculations. They were taken into account in the DM results presented in Sec.~\ref{sec:DMreac}
however.
The main new features of the present calculations with respect to previous works are:
\begin{enumerate}
	\item The anti-nuclei production cross sections used are more tightly constrained by the available 
     data~\cite{DU03}.
	\item  The standard coalescence model for the production cross section of composite particles has been 
     calibrated and validated on the available inclusive production cross section of light antinuclei
     (\db, \tb, \hetb), 
     and an independent microscopic (diagrammatic) coalescence approach was also used.
	\item The inelastic rescattering of the transported particles are taken into account in a more 
     realistic way than in previous works.
	\item  The contribution of the $\bar{p}p\rightarrow\bar{d}X$ reaction cross section was 
     included in the calculated \db flux.
\end{enumerate}

\subsection{Antiprotons and antideuterons}
For the \pb flux, the two source terms are
%
%
%
\begin{eqnarray}
Q^{\rm sec}(T_{\bar{p}})=2\sum_{\rm i=CRs}^{\rm p,He,CNO}\;\sum_{\rm j=ISM}^{\rm
H,He,CNO} \hspace{3.75cm}\nonumber \\
\times 4\pi n_{j}\int_{6\,m_p}^{\infty}
\frac{d\sigma^{\rm i+j}}{dT_{\bar{p}}}(T_{\bar{p}},T_{i})\Phi_{i}(T_{i})dT_{i}\;\;\;\;
\label{eq:5-12}\\
Q^{\rm ter}(T_{\bar{p}})=4\pi. n_{p}\left[2\int_{T_{\bar{p}}}^{\infty}
\frac{d\sigma^{\bar{p}p\rightarrow\bar{p}X}}{dT_{\bar{p}}}(T'_{\bar{p}},T_{\bar{p}})
\Phi_{\bar{p}}(T'_{\bar{p}})dT'_{\bar{p}} \right. \;\;\;\;\;\nonumber \\
\left. -2\sigma^{\bar{p}p\rightarrow\bar{p}X}_{\rm ina}
(T_{\bar{p}})\Phi_{\bar{p}}(T_{\bar{p}})\right]\;\;\;\; \label{eq:5-18}
\end{eqnarray}
where $n_{p}$ is the hydrogen number density in the ISM in cm$^{-3}$, 
$d\sigma^{\rm i+j}/dT_{\bar{p}}$ is the differential \pb production cross section, 
$d\sigma^{\bar{p}p\rightarrow \bar{p}X}/dT_{\bar{p}}$   
is the differential inelastic non-annihilating cross section for \pbs with incident energy 
$T'_{\bar{p}}$ emerging from the collision with an energy $T_{\bar{p}}<T'_{\bar{p}}$, 
and $\sigma^{\bar{p}p\rightarrow\bar{p}X}_{\rm ina}$ is the total inelastic scattering 
$\bar{p}p\rightarrow \bar{p}X$ cross section.
This latter term involves the quantity $\Phi_{\bar{p}}(T_{\bar{p}})$ in the integrand, 
requiring a numerical method to be used to solve Eq.~(\ref{eq:5-11}). Following~\cite{DO01}, 
the tertiary contribution on $H\!e$ is taken into account assuming a mere scaling factor 
$4^{2/3}$ to the corresponding cross section on Hydrogen.
The factor of 2 in front of both terms in \ref{eq:5-12} takes into account the \nb 
(decaying into \pb) production cross section, assuming the \pb and \nb production cross 
sections to be the same. 

For antideuterons, the source terms are similar (with \pb labels changed to \db) with only 
the $pp$, $pHe$ and $H\!e\,p$ incoming channels for the production reaction (threshold 
$16m_p$ for the $pp$ incoming channel) being taken into account (heavier components neglected). 
The new contribution $Q^{\rm ter}$ ($\bar{p}p$ and $\bar{p}He$, threshold $6m_p$) which was 
assumed to be negligible in~\cite{DBDM} was included here. 
In $Q^{\rm sec} (T_{\bar{d}})$ the term $d\sigma^{\rm CRs+ISM}/dT_{\bar{d}}$ was evaluated by 
means of the coalescence models discussed in Sec.~\ref{COALM}.

Note that for the primary fluxes of $p$ and $He$, the parametrization provided by the AMS01 
experiment was used. Since the flux measurements of the BESS \cite{HA04} and AMS01 \cite{AMS01} 
experiments are now compatible, the primary fluxes can then be considered as being a minor source 
of uncertainty of the calculations \cite{DO01}.
\subsubsection{The secondary source term}\label{subsubsec:ter}
\begin{figure*}[htb]          
\begin{center}
\includegraphics[width=\columnwidth,angle=0]{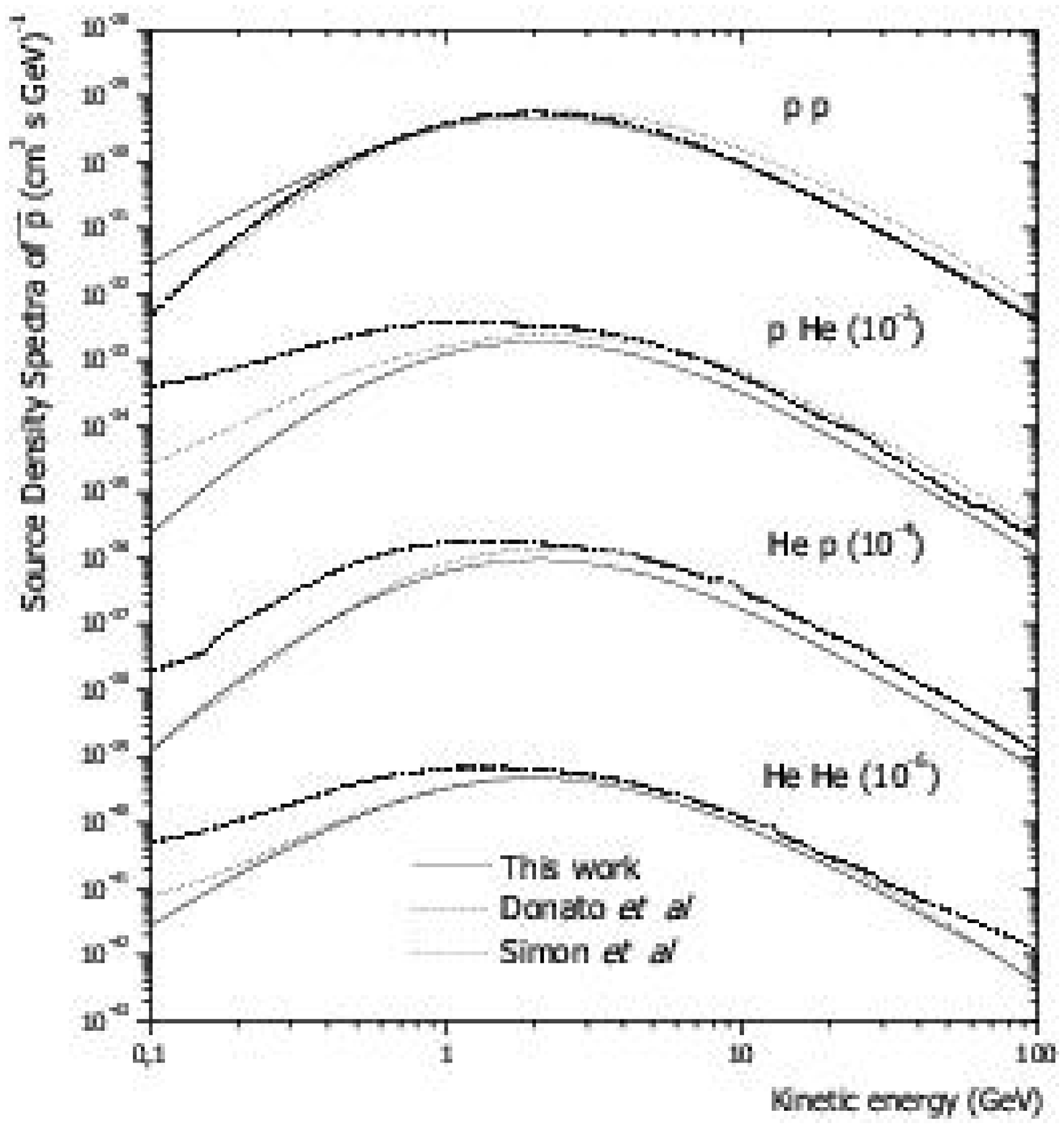}
\includegraphics[width=\columnwidth,angle=0]{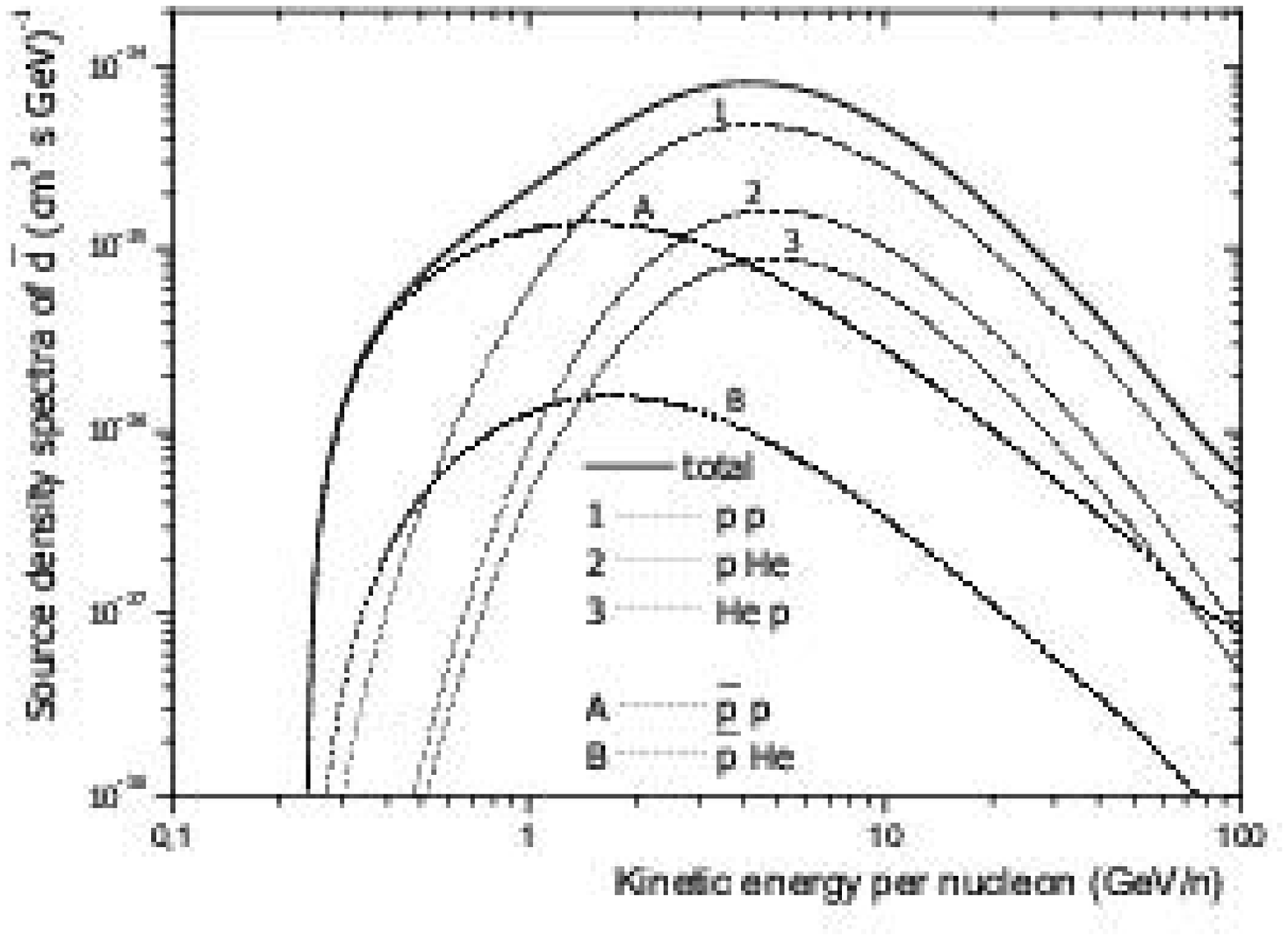}
\vspace{-0.5cm}
\caption{\it\small Source terms for antiproton and antideuteron particles as defined in  
relation~\ref{eq:5-12}.
Left panel:  $Q^{\rm sec}$ for the antiproton production from the present work (solid lines), 
compared with the results from two different papers using the DTUNUC event generator
(\cite{DO01}: dashed lines and \cite{SI98}: dotted lines). Note that the curves
taken from \cite{SI98} (dotted lines) correspond to a slightly different H and He input fluxes
compared with the two other models presented. 
Right panel: $Q^{\rm sec}$ for the antideuteron production for the different channels considered 
in this work. See text for details.
\label{GALpbdbCompos}
}
\end{center}
\end{figure*}
 Fig.~\ref{GALpbdbCompos} shows the various contributions to the source term 
$Q^{\rm sec}(T_{\bar{p}})$ as defined in relation \ref{eq:5-18} (independent of any 
propagation model) for the antiproton (left panel) and antideuteron (right panel) fluxes.

For the antiproton flux, the $pp$ reaction contributes to about 56$\%$ to the $\bar{p}$ 
production, the $pH\!e$ to 24$\%$,  $H\!e\,p$ to 12$\%$ and the $H\!eH\!e$ reaction for 6$\%$. 
The latter has been evaluated by scaling the $pp\rightarrow\bar{p}X$ cross section using 
the same procedure as described in \cite{MO02}. The reactions $p-CNO$ and $CNO-p$ contribute 
to less than $2\%$, while $He-CNO$ and $CNO-He$ have been neglected in account of the very 
low CNO nuclei flux and IS density.
These components are compared on the figure to those obtained in \cite{DO01,SI98} using the 
cross sections calculated using the DTUNUC event generator. At the production peak, for $pp$ 
and $pHe$ reactions, the calculated yields are similar up to 10\%. There is a trend of the 
present calculations to predict smaller cross sections than DTUNUC for particle energies 
above a few GeV. 
In the low energy range $T_{kin}\lesssim$1~GeV, it is seen on the figure that DTUNUC clearly 
predicts a larger \cite{SI98} or even much larger \cite{DO01} target compositness effect than 
in the present work, with a larger low energy cross section calculated for composite targets 
(A$\ge$4) than for the proton target. This difference would deserve a further investigation 
both theoretical and experimental, of the low energy \pb yield in nuclear collisions.
Note however that for the $pp$ induced yield, the present calculations predict a larger flux
at very low energies than the DTUNUC based calculations. It must also be remembered that this 
energy range is marginally within the DTUNUC domain of validity. 

For the antideuterons (right panel), the contributions from the $pp$, $pHe$, and $H\!e\,p$ 
collisions are peaked around 4~GeV/n and the obtained distributions display a similar bell
shape as obtained for the IS $\bar{p}$ flux. This feature can be easily explained 
qualitatively since the calculated distributions of the \db(\pb) fluxes are driven on the 
low energy side by the rapidly raising \db(\pb) production cross section with energy above 
the \db(\pb) production threshold (see \cite{AN73} for example), while the high energy decay 
of the distribution is determined by the rapidly decreasing incident CR proton flux with 
energy, folded with the natural decrease of the high energy production cross section (see 
Fig.\ref{COALSD}).

These dynamical and kinematical characteristics (see subsection~\ref{PPBDB}) are responsible 
for the $\bar{p}p\rightarrow\bar{d}X$ contribution to the flux to be clearly dominant over the 
entire low energy range below 1~GeV/n with a value larger than the value reported in 
\cite{DBDM,DBGLX} by about one order of magnitude (note that the $\bar{p}\,^4\!He\rightarrow\bar{d}X$ 
contribution can be neglected here, as it can be seen by comparing curve A to curve
B in Fig.~\ref{GALpbdbCompos}). At energies above 2~GeV, this flux becomes rapidly negligible.   
\begin{figure*}[htb]          
\begin{center}
\includegraphics[width=\columnwidth,angle=0]{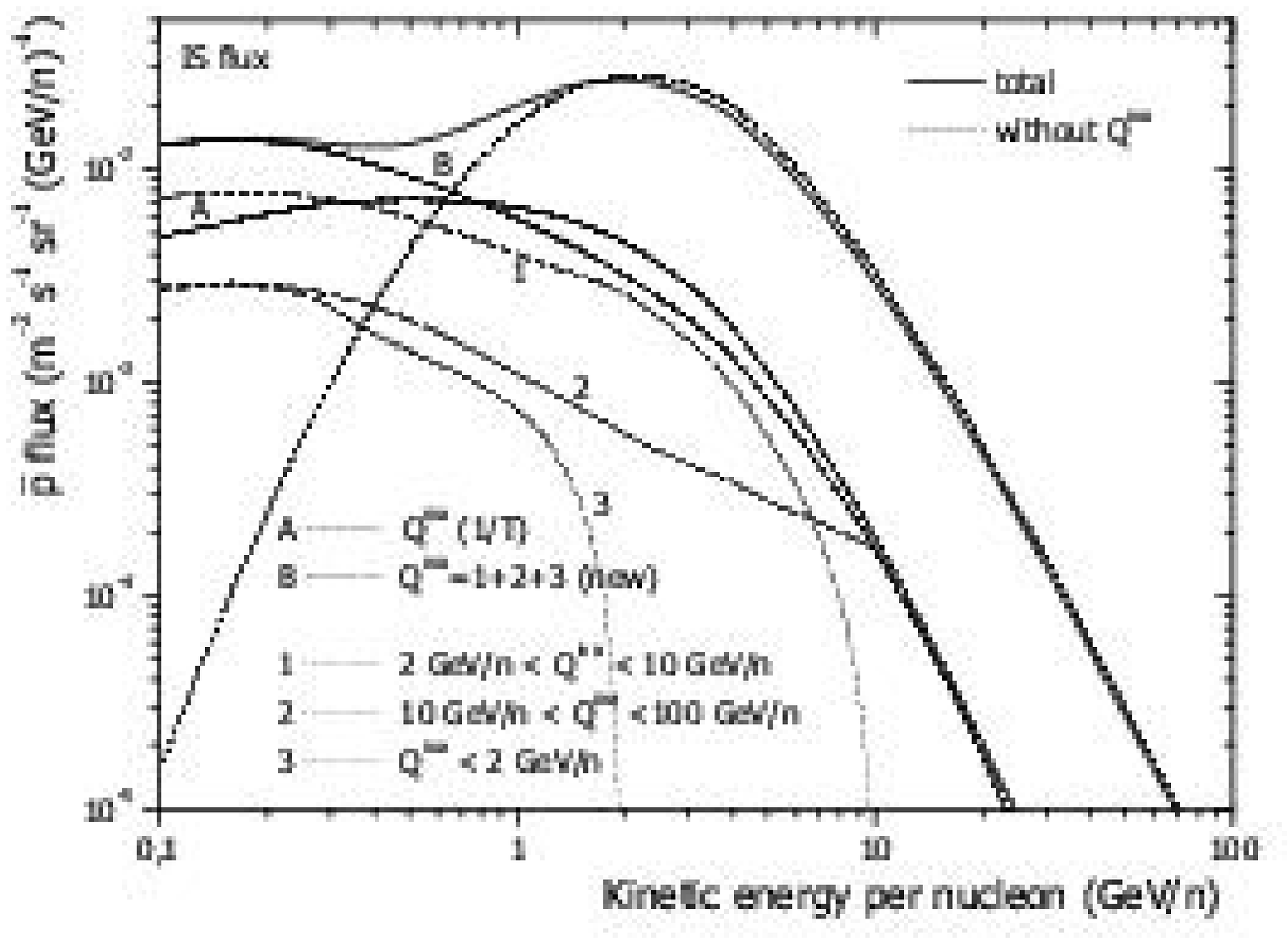}
\includegraphics[width=\columnwidth,angle=0]{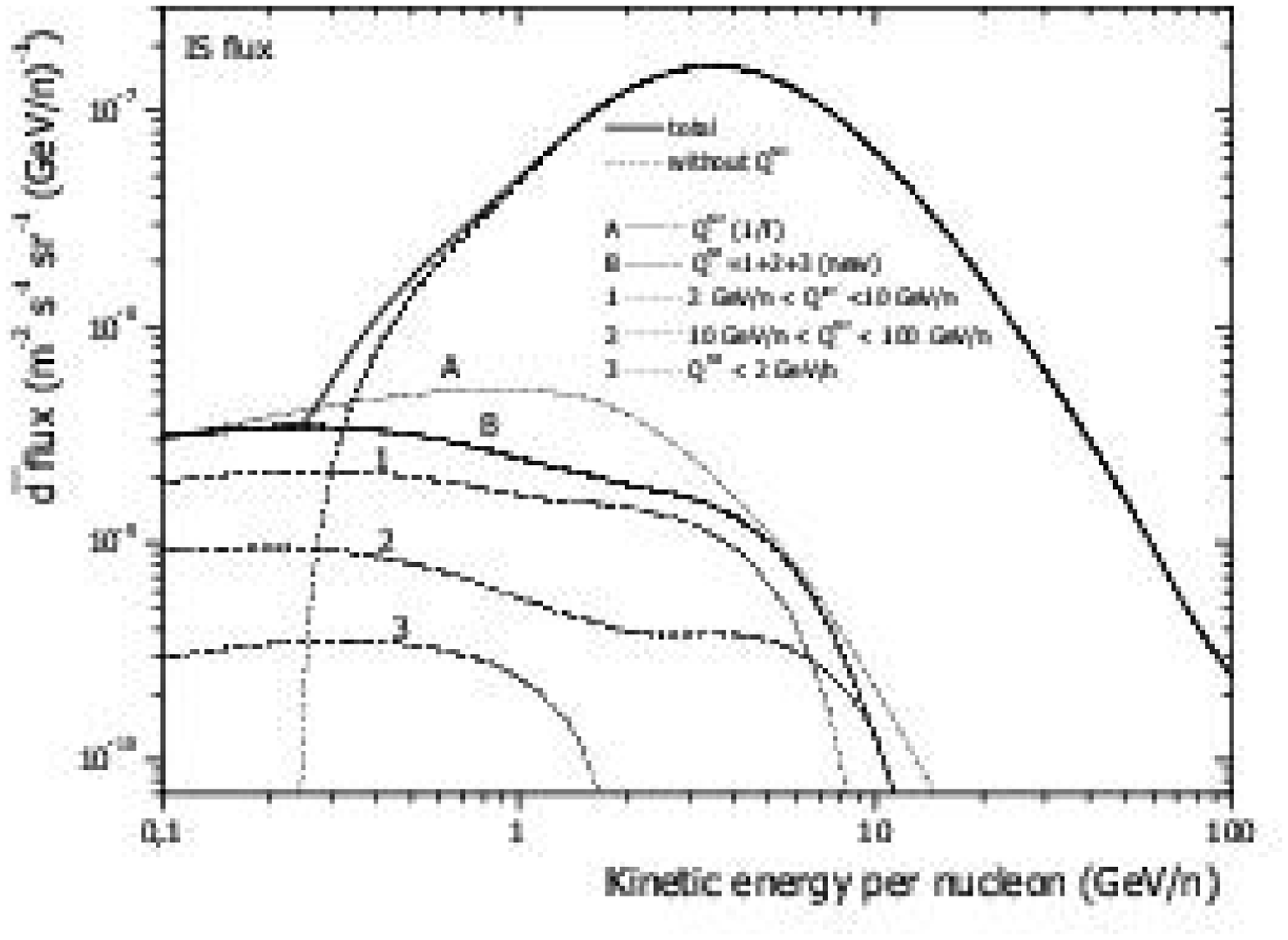}
\vspace{-0.5cm}
\caption{\it\small Comparison of tertiary contribution using the standard $1/T$ 
parametrization (A) and the parametrization used in this work (B) for the \pb (left) and
\db (right) fluxes. The B tertiary component is further decomposed into individual 
contributions from three energy bands (labels 1, 2 and 3). See text for details.
\label{GALpbdbNAR}
}
\end{center}
\end{figure*}

Note that for this channel, the \pb flux was calculated conservatively using the 
parametrization providing the smaller values for the cross section. The range of 
integration for the \db flux calculation has been limited to 100~GeV $\bar{p}$ incident 
energy (while for $pp\rightarrow\bar{p}+X$, the integration was performed up to a few tens 
of TeV $p$ energy to calculate the \pb production), since the \pb flux beyond 100~ GeV
is vanishingly small. The major 
source of error here is due to the uncertainty on $\bar{p}$ flux in the few GeV range. For 
example, a roughly twice larger $\bar{d}$ flux could be produced in this channel by taking 
the mean values of the experimental data points for this flux, rather than the $\bar{p}$ 
flux obtained from the \pb data fits. 
For the other channels, the uncertainties are related to those for the \pb production, 
combined to the contribution induced by the coalescence model. It must be emphasized however 
that, although the procedure used to evaluate this cross section is based on the underlying 
physics, the error induced by the approximations made in this approach cannot be accurately
evaluated.

In conclusion of this section, it has been shown that the secondary $\bar{p}p\rightarrow\bar{d}X$
term does contribute significantly to the IS galactic \db flux, and that it is even dominant 
in the low kinetic energy $T_{\bar{d}}\lesssim$~1.5~GeV/nucleon range.

\subsubsection{The tertiary source term}\label{TERTIARY}
The only tertiaries considered here are those created by the non-annihilating inelastic 
scattering of secondaries on the ISM (NAR process).This contribution was first included 
for \pbs in \cite{BE99b}.

The sharp kink observed in the low energy \db spectrum in Fig.~\ref{GALpbdbNAR} right, can 
be better understood by examining how inelastically scattered antiprotons or antideuterons 
are redistributed, since the redistribution mechanism may have significantly different 
effects on the respective shapes and intensities for the \pb and \db flux.
Since the $Q^{\rm ter}$ term in Eq.~(\ref{eq:5-18}) has to be integrated numerically, the 
same iterative procedure as proposed in~\cite{DO01} was used. 
For antiprotons, a few iterations are required, while only one single iteration is needed 
for antideuterons. This is simply due to the much larger relative inelastic (NAR) cross 
section compared to total cross section for \pb\!$p$ than for \db$\!p$ collisions (see 
section~\ref{NAR} and appendix), which makes the numerical convergence slower in the 
former case.
To better understand the details of the transfer process to lower energy, the tertiary 
contributions from three kinetic energy bands (T$<$2~GeV, 2$<$T$<$10~GeV, 10$<$T$<$100~GeV) 
are displayed separately on Fig.~\ref{GALpbdbNAR}. The corresponding curves were obtained by 
first evaluating the equilibrium spectrum, and then computing from this spectrum the tertiary 
yield from the chosen energy bands.

For the antiprotons (Fig.~\ref{GALpbdbNAR}, left panel), the NAR cross section is large and 
the low energy tail is largely replenished. Several iterations are required to obtain the 
equilibrium flux allowing the second order NAR contribution to be significant, 
the first iteration replenishing both the medium and low energy bands, the former of 
those replenishing the low energies in the next iteration step.
 
For the antideuterons, the NAR cross section is much smaller than for $\bar{p}$s. The first 
consequence is that the tertiary component is almost two orders of magnitude smaller than 
the peak value of the secondary component (while it is less than one order of magnitude 
smaller for $\bar{p}$s). One single iteration is thus required in the process and the 
contribution of second order interactions is negligible. This explains the sharp upturn in 
the \db flux that is not seen in the \pb spectrum. The secondary flux drops rapidly at low 
energy and the tertiary and secondary components become comparable only at 
$\approx$~300~MeV/n.
In this case the NAR process doesn't accumulate particles at low energies as efficiently as
in the antiproton case because of the much smaller NAR cross section (compare the two curves 
labelled 3 in the left and right panels). 

The effects of the NAR spectrum on the \pb and \db distributions are also illustrated on the 
figure, with the curves labelled A and B corresponding to the parametrization used in 
\cite{FI74} and here (see section~\ref{NARpractical}), respectively. Although for \dbs the 
results are hardly different at the lowest energies, for \pbs, the more realistic inelastic 
spectrum leads to a low energy NAR flux larger by almost a factor of 2 than the other option.
\begin{figure*}[htb]          
\begin{center}
\includegraphics[width=\columnwidth,angle=0]{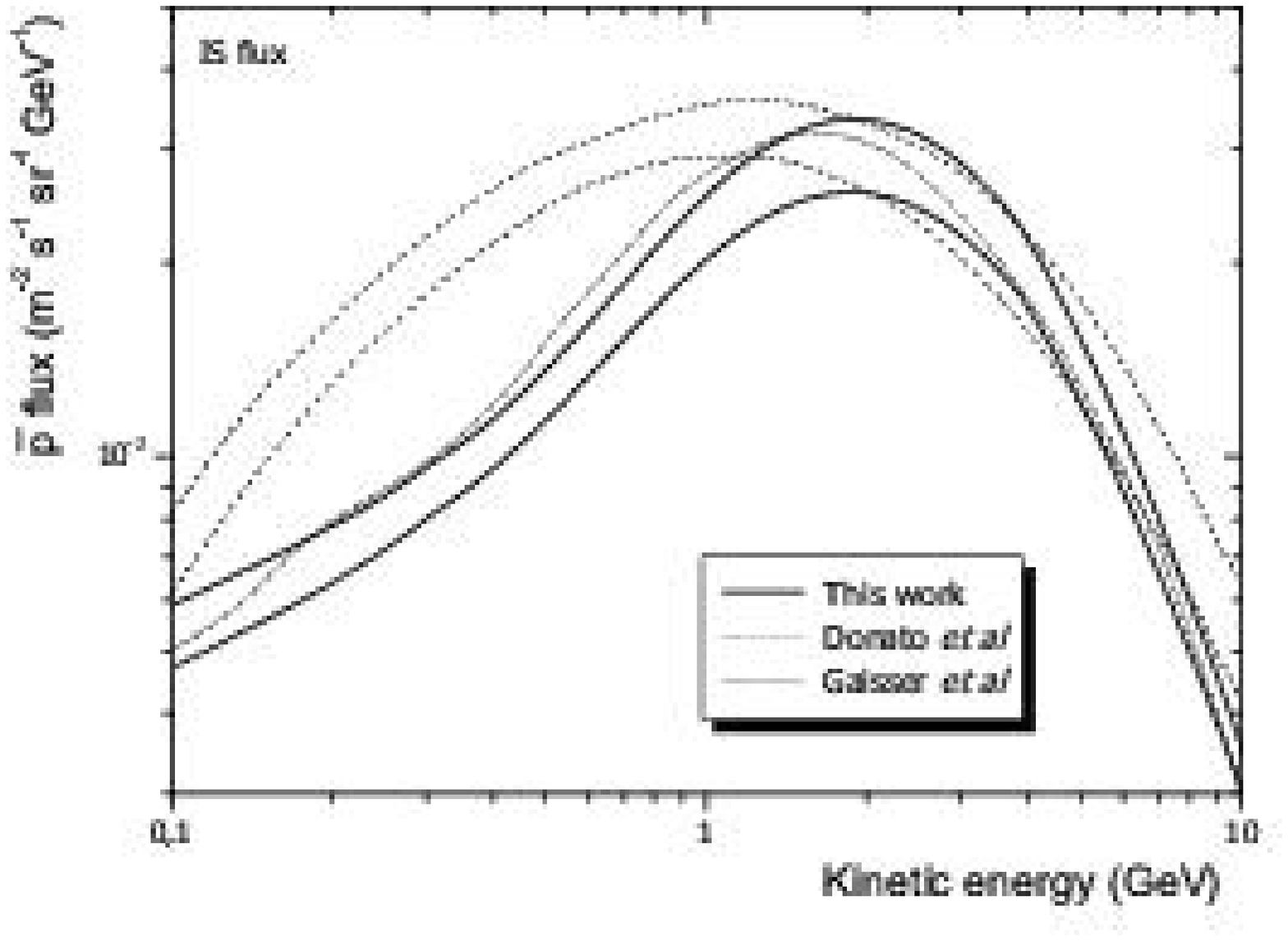}
\includegraphics[width=\columnwidth,angle=0]{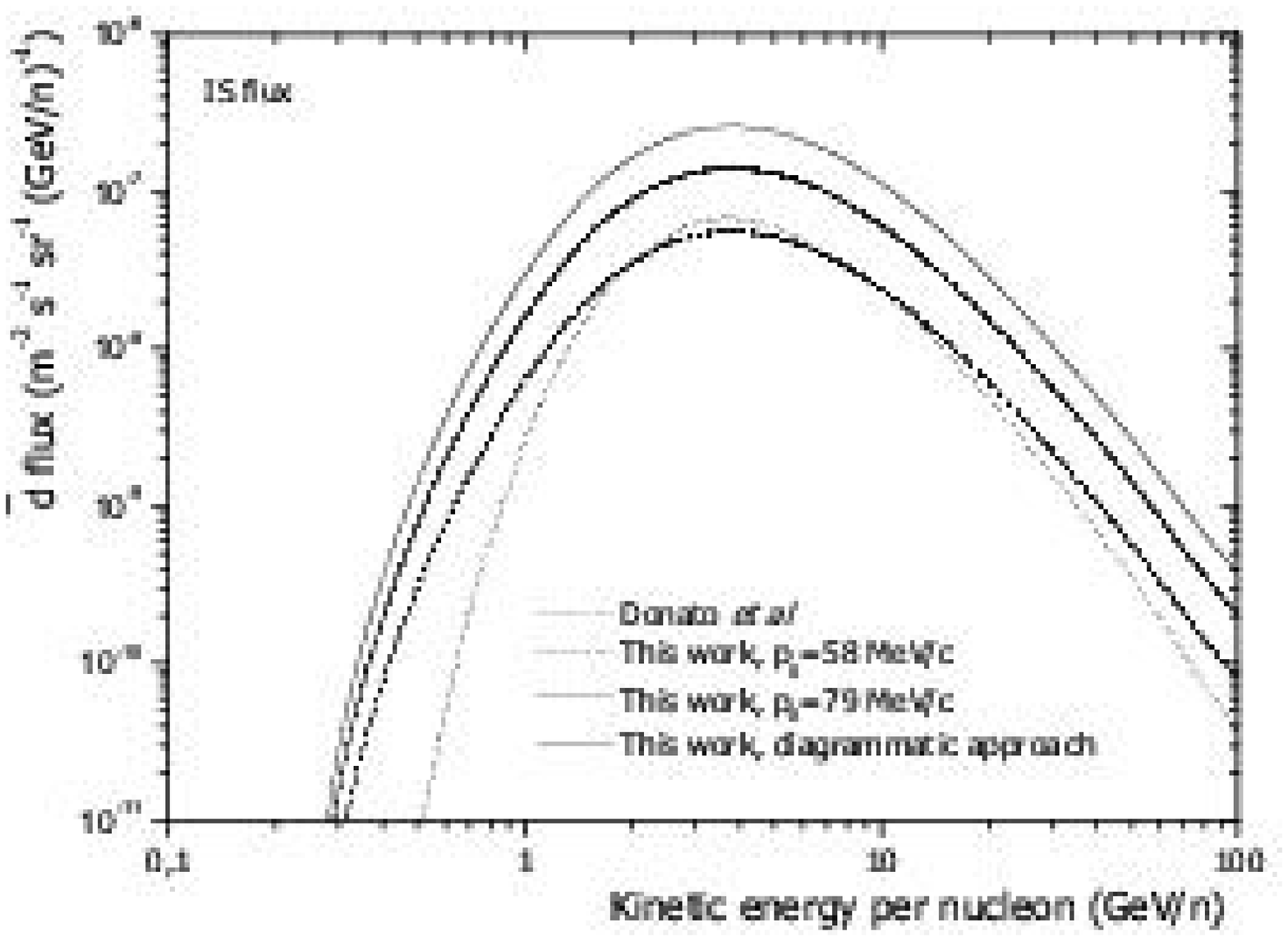}
\vspace{-0.5cm}
\caption{\it\small Partial calulations of Galactic antiproton (left panel) and antideuteron 
(right panel) interstellar fluxes for comparison purpose. 
Left panel : The present calculations (solid lines) are compared with 
those from \cite{DO01} (diffusion model with reacceleration, dashed lines) and from 
\cite{GA99} (leaky box model, dash-dotted line). 
The two curves from the present work (using standard NAR), obtained for the two different 
parametrizations of the production cross section, as well as those from \cite{DO01}, indicate 
the level of uncertainty of the results due to hadronic cross sections. 
Right panel (no tertiaries): Present results ($pp$, $pHe$ and $H\!e\,p$ contributions only) 
using the standard coalescence model (fitted to the \db data) with $p_0=79$~MeV (dash-dotted 
line) and the diagrammatic approach of coalescence (solid curve).
The two lower curves show the present calculations using the same coalescence parameter 
$p_0=58$~MeV (dashed line) as in \cite{DBDM} (diffusion model, no reacceleration, dotted 
line). See text for details.
\label{GALpbdbCompar}
}
\end{center}
\end{figure*}
\begin{figure*}[htb]          
\begin{center}
\includegraphics[width=\columnwidth,angle=0]{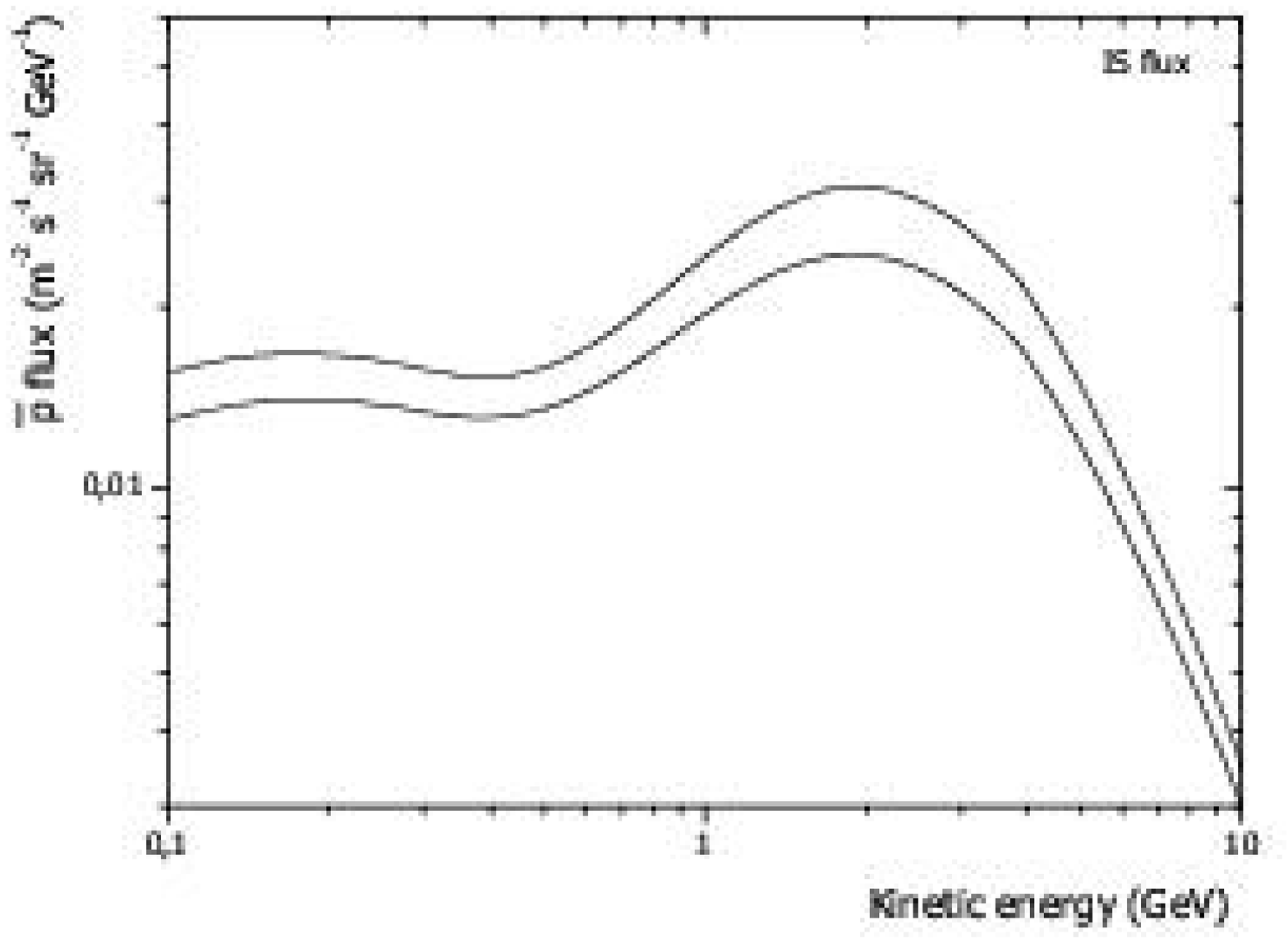}
\includegraphics[width=\columnwidth,angle=0]{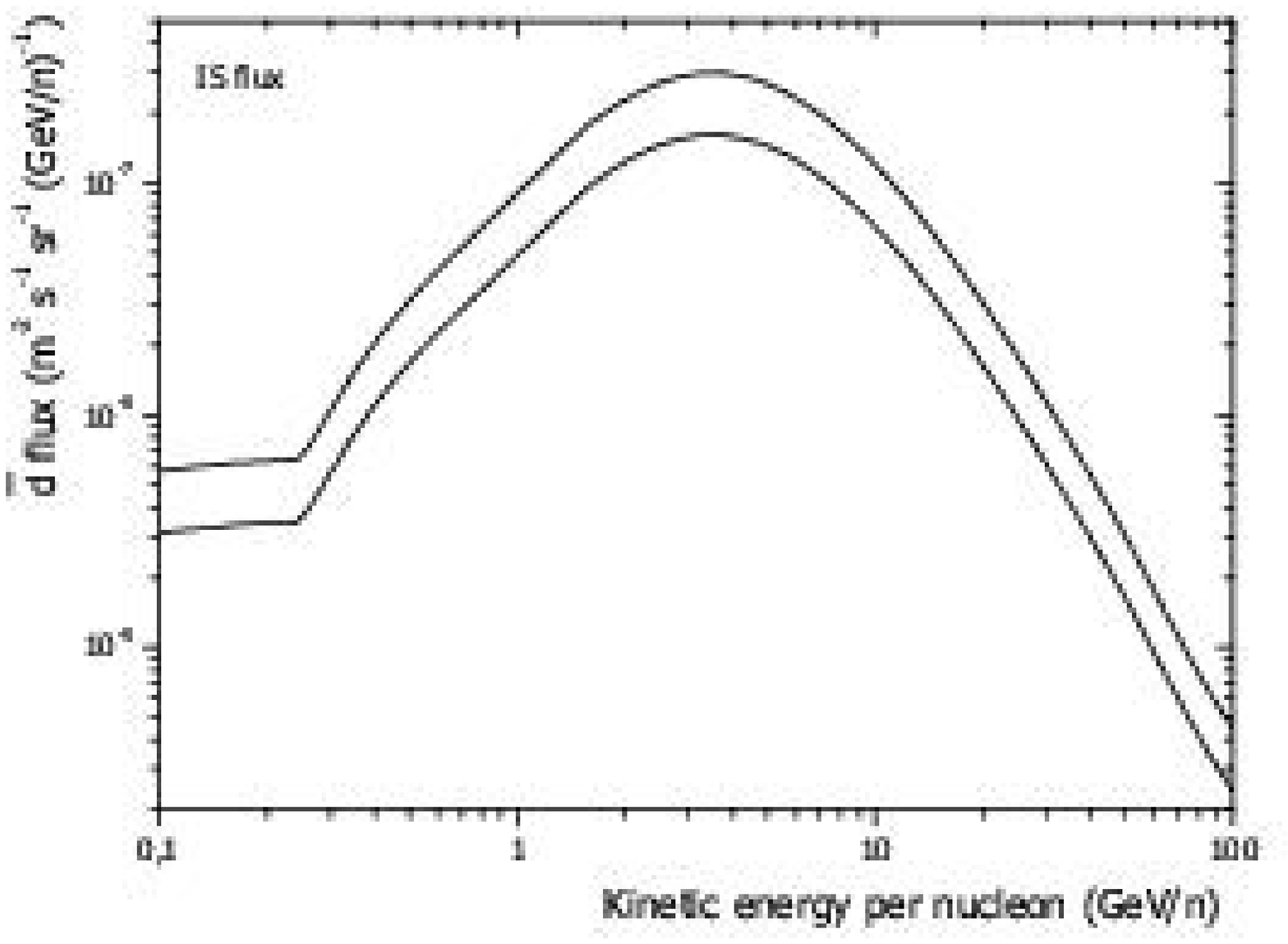}
\vspace{-0.5cm}
\caption{\it\small Left panel: IS \pb flux including all contributions, the two lines showing 
the uncertainty due to nuclear cross sections (see text). Right panel:  \db flux for the 
standard coalescence model (reference calculation, lower curve) and for the microscopic 
coalescence model.
\label{GALpbdbIS} }
\end{center}
\end{figure*}
\subsubsection{Discussion of the interstellar fluxes}\label{ModulFlux}
\paragraph{Comparison to previous published fluxes}

\

Left Fig.~\ref{GALpbdbCompar} shows the results from the present work for the unmodulated 
antiproton fluxes compared with some previously published results. For the present 
calculations (full lines), the two curves correspond to the two different parametrizations 
used for the \pb production reaction in $pp$ collisions \cite{DU03}. The present results are 
close to those from \cite{GA99} (dashed line). Both rely on better fits to the data than in 
\cite{TA83}. The latter, used in the present approach, would give a flux standing roughly 
midway between the two full lines for \pb energies above 1~GeV. The so defined range can 
thus be considered as the uncertainty due to the production cross section. 

In the results of the diffusion model \cite{DO01}, the flux is somewhat higher at high 
energy, while at low energy the effect of reacceleration is responsible for the much larger 
predicted flux (note the enlarged vertical scale in Fig.~\ref{GALpbdbCompar}). However, it 
will be seen below that the two results are not so different after modulation. Note that as 
the parameters used for the present calculations provide the correct magnitude for the 
calculated fluxes (see left Fig.~\ref{GALpbdbMod}), they can be used confidently to evaluate 
the galactic fluxes for other antinuclei.

Right Fig.~\ref{GALpbdbCompar} shows the calculated antideuteron flux (dotted line)
evaluated as in ref~\cite{DBDM} (dashed line), i.e., including neither the tertiary term (NAR) 
nor the $\bar{p}p$ included and using a coalescence parameter $p_0=58$~MeV.
Note that the p and He galactic fluxes used in \cite{DBDM} were slightly 
different however from the more recent measurements used here. Note also that in \cite{DBDM}, 
a diffusion model without reacceleration was used. The two calculations appear to be 
compatible. At low energy, the difference observed in the antideuteron flux from the two 
approaches originates from:
\begin{enumerate}
	\item The difference between the low energy \pb production cross section used here and that 
from  \cite{TA83} (dominant effect), and
	\item The kinematics of the \pb production near the threshold, also modelled differently here 
than in \cite{DO01}, providing a larger antiproton - and thus antideuteron~- cross section, 
over this range.
\end{enumerate}
The upper two curves (solid lines) on the figure, correspond to the standard model (reference
calculation, lower curve) and to the microscopic model calculations discussed in Sec.~\ref{DBARCOAL}
(upper).

\paragraph{Full calculation}

\

Fig.~\ref{GALpbdbIS} displays the full calculation results for the IS fluxes obtained in 
this work. The left panel shows the two calculations showing the uncertainty on the $pp$ 
production channel induced by the two parametrizations used (dedicated $pp$ (II), and $pA$ 
with A=1 (I), see section~\ref{PBPROD} and \cite{DU03}). 

It was seen in sections~\ref{DBARCOAL} and in \ref{subsubsec:ter} that the uncertainty for 
the other channels that contribute to almost half of the flux, is similar, their contributions 
having been evaluated using widely proven calculation methods based on existing experimental 
data. Comparing with the results obtained using the DTUNUC generator, large differences have 
been found for the various individual channels contributions, although a rough agreement for 
the overall secondary production has been obtained. 

For the antideuteron flux (Fig.~\ref{GALpbdbIS}, right panel), the lower curve shows the 
reference results obtained with the standard coalescence model (based on a fit of \db 
data) and including all the secondary and tertiary components discussed previously. 
The overall uncertainty associated to the reference calculation is of the same order of 
magnitude as for the \pb flux, plus the coalescence model contribution to be added (see 
section~\ref{DBARCOAL}). It is estimated to be better than a factor of 2. The upper curve 
on the figure shows the microscopic coalescence model calculation (which overestimates the 
experimental data on the average). 

At low energy, the $\bar{p}p\rightarrow$\db$X$ reaction can be estimated to be correct to 
within a factor of 2 regarding the accuracy of the inclusive $\bar{p}$ available data. 
The tertiary term corresponds to a pretty conservative lower limit since, based on the 
evaluation method used (see section~\ref{NARpractical}), the upper limit being less than a 
factor of 2.
From this point of view, some direct measurements of the $\bar{p}p\rightarrow$\db$X$ 
production reaction, would be extremely useful to eliminate the uncertainties corresponding 
to the approximations made here.
\begin{figure}[htbp]          
\begin{center}
\includegraphics[width=\columnwidth,angle=0]{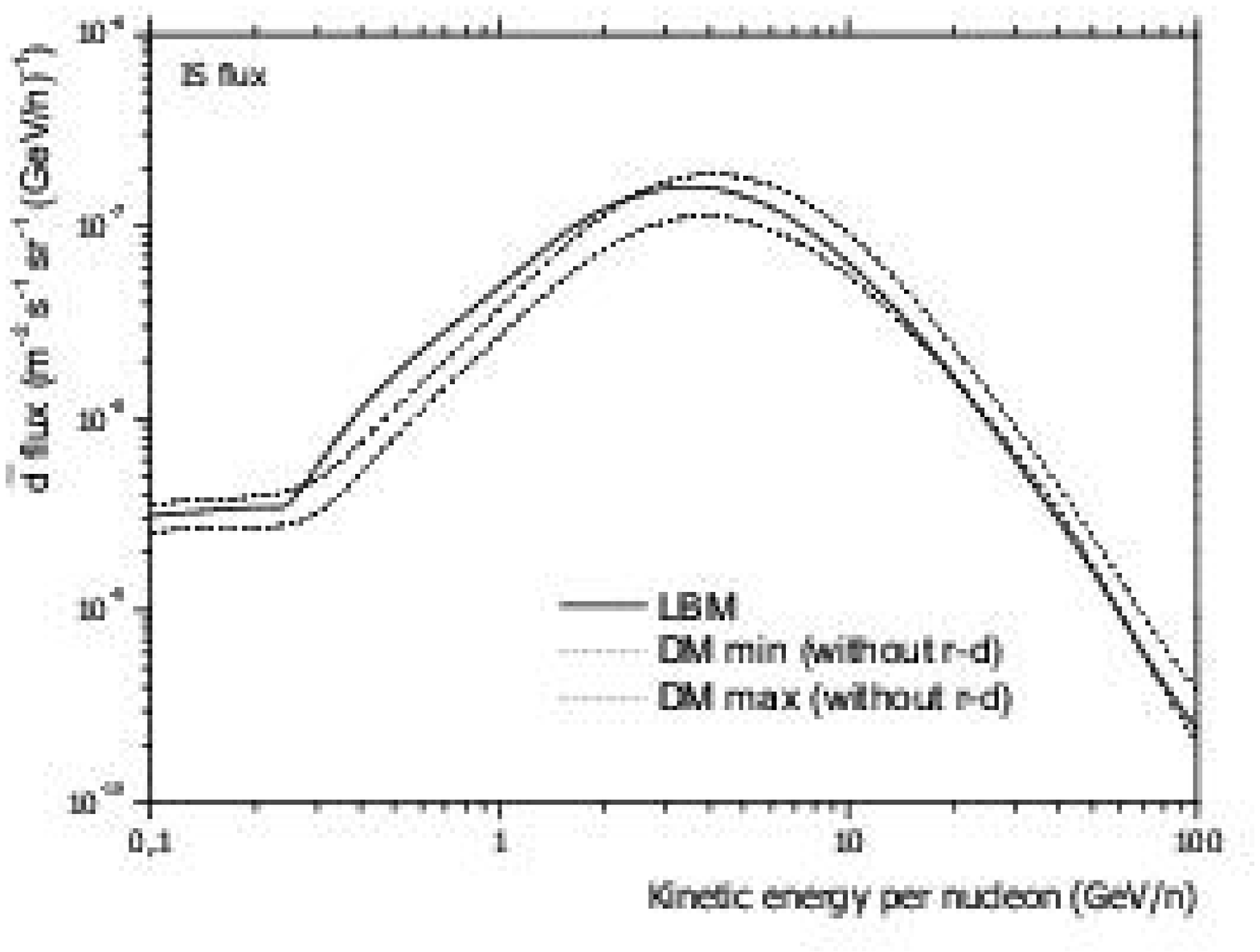}
\includegraphics[width=\columnwidth,angle=0]{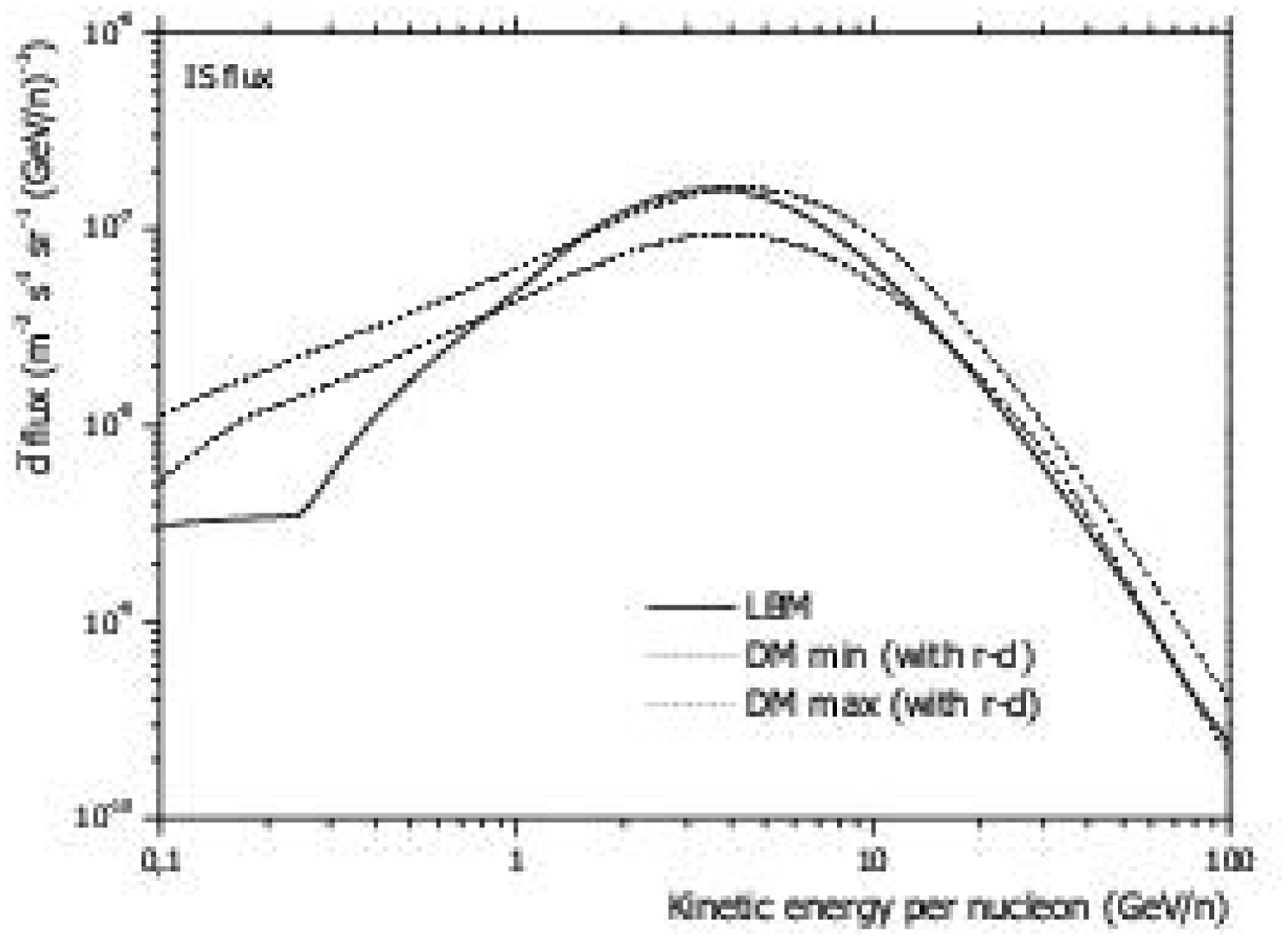}
\includegraphics[width=\columnwidth,angle=0]{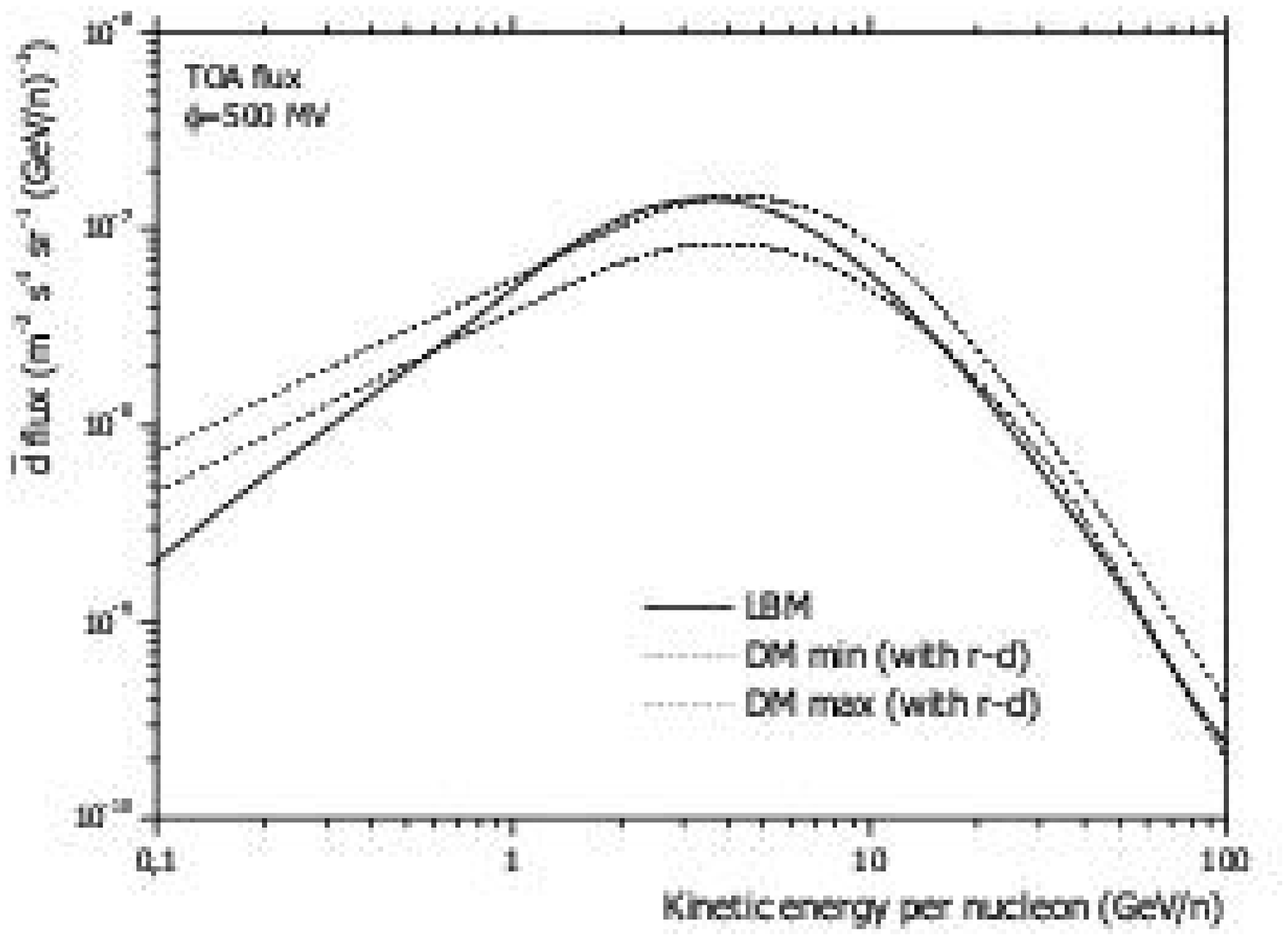}
\vspace{-0.5cm} 
\caption{Comparison of the present IS (unmodulated) \db flux calculations (solid line) with 
the diffusion model results (dashed line), without (Top) and with (Middle) reacceleration 
term included (see \cite{MA01}), and with modulation and reacceleration included (Bottom). 
The two lines for the DM results on the three panels correspond to the same estimated range 
of propagation uncertainty as evaluated in \cite{MA01} and used for \pb in \cite{DO01}. 
\label{ISDMRD}
}
\end{center}
\end{figure}
         \paragraph{Comparison with the diffusion model}
				 \label{sec:DMreac}

\

Fig.~\ref{ISDMRD} shows the diffusion model calculations for the \db flux, 
with the same common inputs as in the present calculations, i.e., same $p$ and $He$ galactic 
flux and hadronic cross sections, including the $\bar{p}p\rightarrow$\db$X$ contribution. 
The top panel shows the results without diffusive reacceleration for the DM calculations. 
The agreement between the two results is good, except in the 500~MeV/n range where the 
contribution of the $\bar{p}p\rightarrow$\db$X$ contribution is more salient in the present 
calculations than in the DM approach, making the predicted flux larger by a factor of about 
two.
The middle panel shows the DM calculations including the diffusive reacceleration term 
compared with the same present results as above. The observed effect of the reacceleration 
on the spectrum is similar to the effect of the solar modulation (compare with curve 1 in 
Fig.~\ref{GALpbdbMod} right), making the distribution at low energies below 1~GeV/n 
power-law shaped. The bottom panel shows the effects of the combined reacceleration term and 
solar modulation for the DM model, compared with the present calculation including the solar 
modulation effects. In this case, the general shape of the results are similar, the very low 
energy flux being predicted larger from the DM calculations than from the present LBM by a 
factor of 2 to 4.
\begin{figure*}[htb]          
\begin{center}
\includegraphics[width=\columnwidth,angle=0]{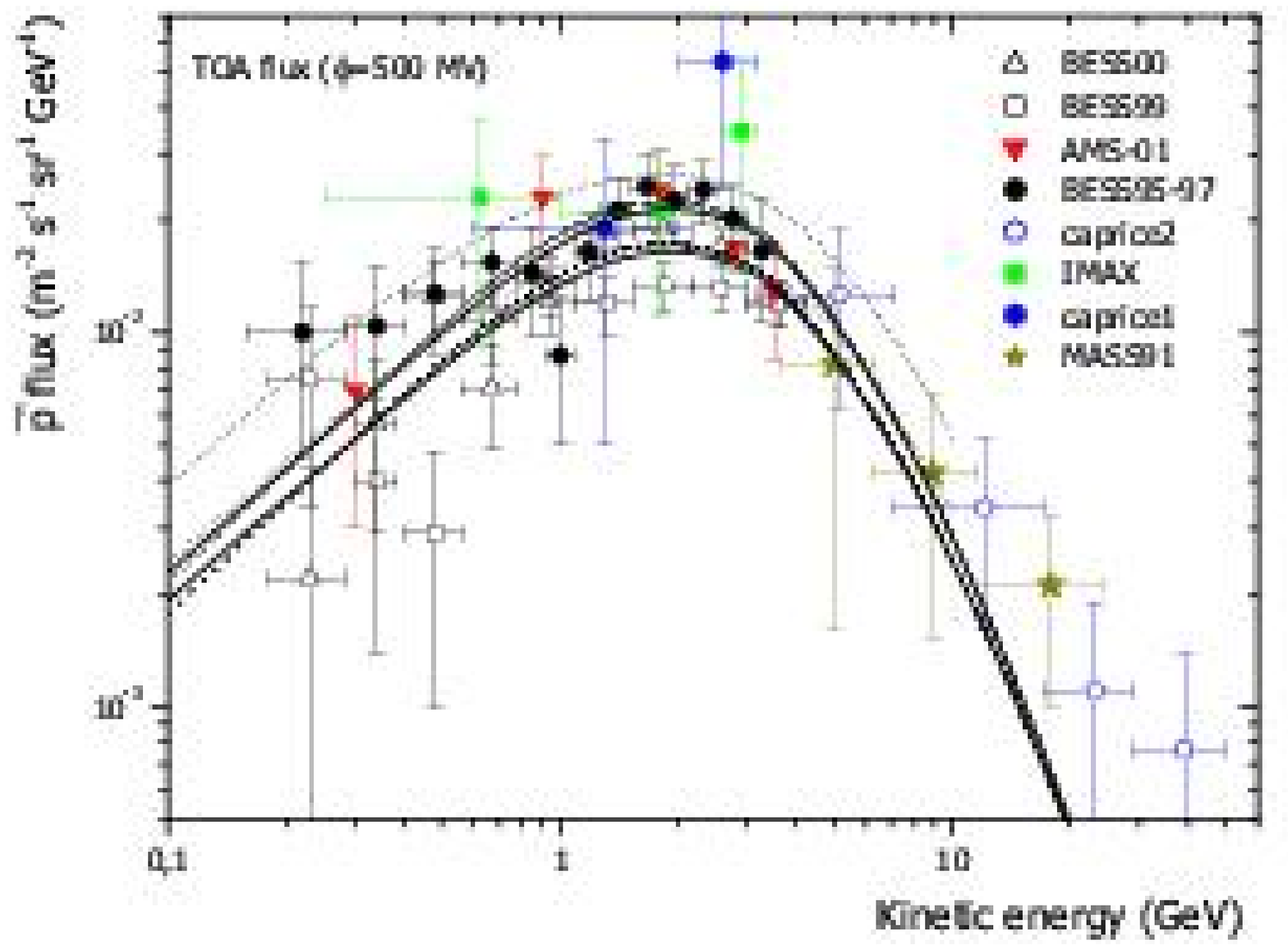}
\includegraphics[width=\columnwidth,angle=0]{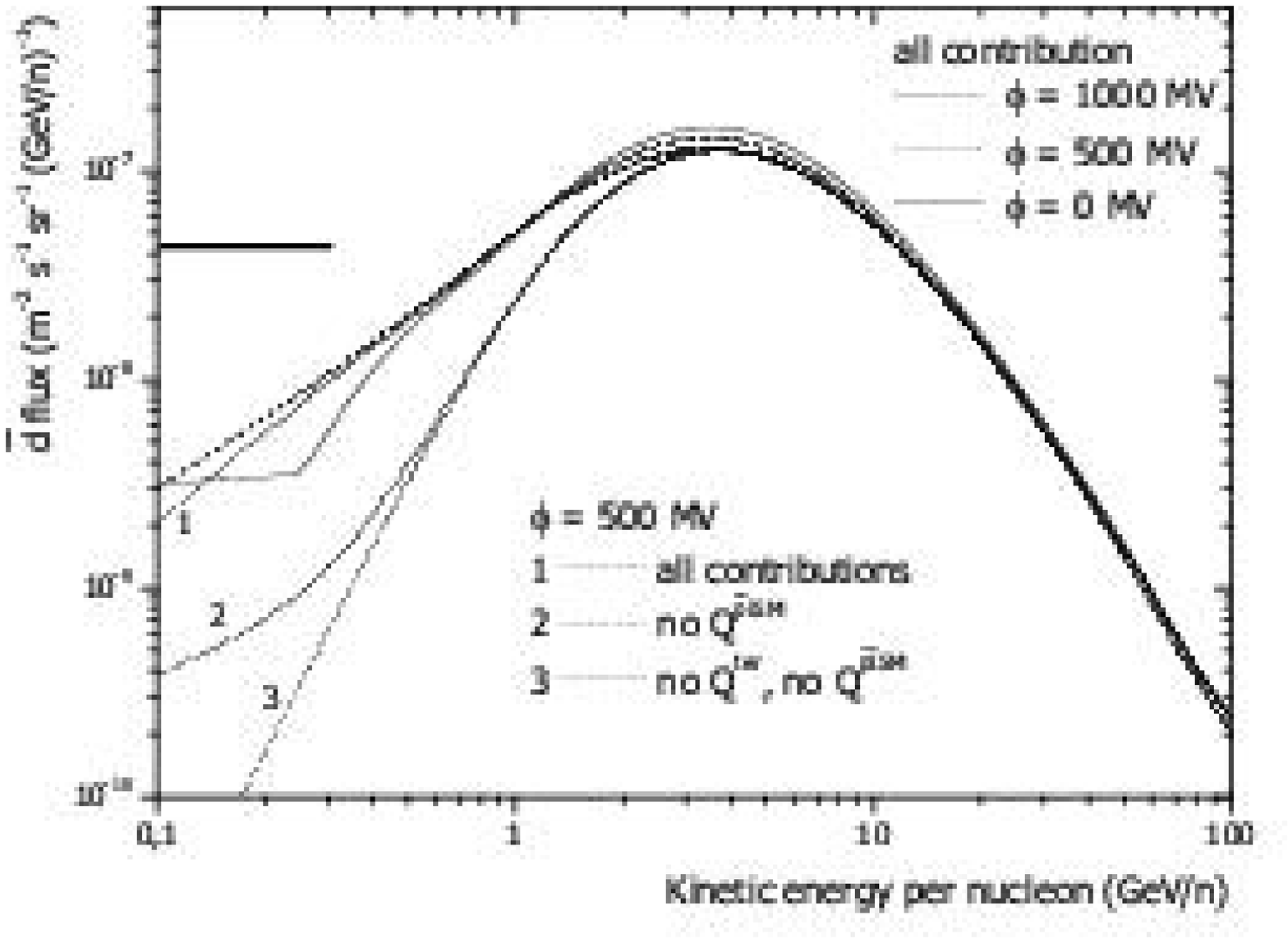}
\vspace{-0.5cm} \caption{\it\small Modulated antiproton fluxes.
Left panel: Present calculations, with the two linetypes corresponding to the single 
$pp\rightarrow \bar{p}X$ reaction, without (solid lines) and 
with (dashed lines) the NAR contribution. They are compared to the DTUNUC calculations 
(dotted lines), the two curves for the latter giving the overall production uncertainties as 
evaluated in \cite{DO01} (see text for details). 
Note that the differences due to reacceleration, observed in the IS fluxes, are completely 
smoothened by the solar modulation effects.
Right panel: Same calculations with the curves labelled 1-3 giving the fluxes obtained, for 
the full calculation (label 1), with the contribution of the new ingredients $Q^{\bar{p}ISM}$ 
switched off (label 2), and with both $Q^{\bar{p}ISM}$ and the NAR $Q^{ter}$ contribution 
switched off (label 3), for the same modulation level ($\phi$=500 MV). The (typical) level 
of primary fluxes evaluated by various authors is shown by the thick solid straight line on 
the left \cite{DBDM,DU04} (see text).
\label{GALpbdbMod}
}
\end{center}
\end{figure*}
\subsubsection{Solar modulation effects}
It has been seen in the previous sections that quite significant differences between the 
antimatter fluxes are predicted in the astrophysically sensitive low energy range of 
the studied particles. In this section, the solar modulation effects on the calculated flux 
are incorporated and their influence on the various components of the calculations together 
with the resulting overall accuracy are discussed.

The situation for the low energy spectrum is drastically different for the modulated fluxes, 
i.e. at TOA, than it was before for the IS spectra. 
Fig.~\ref{GALpbdbMod} left shows the antiprotons flux using the standard, i.e. flat spectrum,
NAR cross section (solid lines) and the parametrization \cite{AN67} used here (dashed 
lines) for a solar modulation parameter $\phi=500$~MV. 
Whereas, these two calculations produced very different IS 
fluxes (see left panel on Fig.~\ref{GALpbdbNAR}), they are hardly different once modulated. 
This is not surprising since the effects of the solar modulation are well known to occult 
the low energy range of the IS spectrum. This effect results in the present calculation in 
improving the agreement (up to a sound normalization factor however) with the results of the 
diffusion model with reacceleration (dotted lines) from \cite{DO01}.
For each case the two bands correspond as before to the uncertainties associated to the 
hadronic cross sections involved: for the present calculations (solid lines) they include a) For the $pp$ 
incoming channel, the uncertainties related to the I and II parametrizations; b) For $pA$ 
($A>$1) systems, the uncertainties of the fit to the data as quoted above (see \cite{DU03}); 
c) For $AA$ ($A>$1) systems those related to the (Glauber) calculation procedure and its 
ingredients \cite{DU04}. 
In \cite{DO01} the uncertainty band was evaluated from the DTUNUC uncertainties 
with all the contributions of the relevant $AA$ collision systems considered, combined 
together.

The \db flux naturally suffers the same source of uncertainty due to the hadronic cross 
sections as for the \pb flux.
Fig.~\ref{GALpbdbMod} right illustrates the status of the various components of the \db flux. 
The three curves labelled 1, 2 and 3, show the relative contribution of the various source 
terms for the same modulation parameter $\phi=500$~MV. As it could be expected from 
Fig.~\ref{GALpbdbCompos}, the \pb induced \db production ($Q^{\bar{p}ISM}$) dominates over 
the \db rescattering (NAR) contribution ($Q^{ter}$) at energies smaller than about 500~MeV/n, 
where the latter is clearly negligible. This is even more true at higher modulation level 
(compare solid line, dotted line 1, and dashed line). 

Finally, the thick solid horizontal line on the figure shows a lower limit for the estimates 
of primary source - supersymmetric particles annihilation (SUSY) or PBH evaporation - 
contributions which fluxes in both cases drop rapidly to negligibly small values above 1~GeV/n.
Some of these predictions~\cite{DBDM,DBPBH} give fluxes that could be one order of magnitude 
larger than the quoted limit. Hence, this confirms that the antideuteron signal is a 
probably good tracer to look for SUSY, likely better than antiprotons, provided that \dbs 
can be discriminated from \pbs and electrons at the appropriate level of selectivity, i.e., 
with a rejection power of the order of 10$^5$, in the forthcoming experiments, which is 
quite an instrumental challenge.
            \subsubsection{Uncertainties on the calculated fluxes}

In summary, in the context of using the \pb flux to derive constraints on new (astro)physics 
(primary exotic component), it can be considered that the NAR cross section is now evaluated 
with an improved accuracy, even if this accuracy is clearly not a major issue here since it 
has been seen that for rigidity R$\lesssim$~0.5~GV of the IS flux, all differences are swept 
away by the solar modulation effects.
On the other hand, it must be emphasized that the significant differences observed between
the results obtained from data fits, and used here for the production cross sections,
and those obtained from the DTUNUC predictions (for $AA$ collision systems), would certainly
deserve a more complete investigation sinceit is a major source of uncertainty for the flux 
calculations.

Below 0.5-1~GeV/n, the contribution of the $\bar{p}p$ induced \db flux which suffers many 
flaws as emphasized in~\ref{subsubsec:ter}, is probably large. Within and below this region 
of energy, several sources of large uncertainties due to, in decreasing order of magnitude, 
hadronic cross sections, solar modulation, and propagation, have been seen to exist. 
These would dramatically increase the difficulty of reliable precision astrophysics measurements 
over this energy range. Above 1~GeV/n, the dominant source of uncertainties remains the hadronic 
cross sections.
  \subsection{Mass 3 and 4 antinuclei fluxes}\label{MASS34G}
The same calculations have been performed for the propagated flux of mass 3 and 4 
antinuclei using the same coalescence model, with $p_0=78$~MeV (note that the diagrammatic 
approach used for the antideuteron case cannot be applied for A=4), and the same hadronic 
cross sections. Unlike antideuterons for which the $\bar{p}p$ reaction provides a sizeable 
flux, similar reactions (e.g., $\bar{d}p\rightarrow \bar{d}X$) are expected to lead to 
negligible contributions for heavier antinuclei production. This is because of the expected 
smallness of the corresponding cross sections, due to the large energy-momentum transfer 
required to produce antinucleon pairs which would breakup the incident (anti)deuterons.

For the tertiary flux of mass 3 antinuclei, the NAR cross section would be large since the 
restriction to isoscalar transitions discussed for the \db projectile would not apply 
(while it would for \heqb), and a similar procedure as used before for \dbs could be 
applied. The exercise would be irrelevant however, in account of the smallness of the 
expected coalescence cross sections and of the corresponding NAR flux. 

The uncertainties associated to these calculations are driven mainly by the uncertainties 
on the hadronic cross sections, boosted by the coalescence model exponentiation. 
For the \hetb flux, it is estimated that the calculations are accurate within roughly 
one order of magnitude, which is a good enough level of accuracy for the purpose of this 
work.
  \subsection{Summary for the calculated galactic flux}

\begin{figure}[htb]          
\begin{center}
\includegraphics[width=\columnwidth,angle=0]{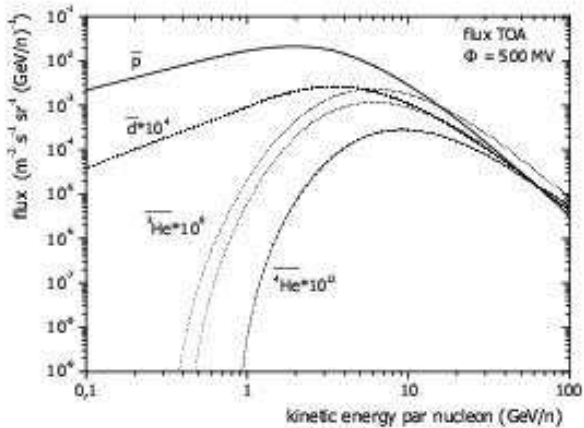}
\caption{\it\small Galactic flux for \pb (solid line), \db (dashed), \hetb (dash-dotted), and
\heqb (dotted) antimatter particles. The lower (resp. upper) dashed line correponds to the 
case where the $^3H$ production is not taken (resp. taken) into account (see text for 
details).
\label{GAL34F}}
\end{center}
\end{figure}

The spectral distributions of the calculated galactic fluxes for mass 1 to 4 antinuclei 
are displayed on Fig.~\ref{GAL34F} (full calculation, standard coalescence model).
The calculated \db,\hetb,\heqb flux have been multiplied by 10$^{4}$, 10$^{8}$, 10$^{12}$ 
respectively for presentation purpose. These fluxes are significantly higher than those 
derived in~\cite{DBGLX}, the difference being mainly due to the larger value of the 
coalescence momentum derived in the present work, the other smaller differences having been 
discussed in the text. The roughly twice larger \db flux between the old and the new 
coalescence momentum translates into a factor 4 for \hetb and 8 for \heqb. The upper of the 
two dashed lines for \hetb includes the addition of the \tb flux, this latter nucleus 
decaying into \hetb with a half life of about 12 years.
%
\section{Atmospheric production and propagation}\label{ATMO}
A CR particle crosses a sizeable amount of the earth atmosphere before being detected by a 
balloon borne running experiment. The grammage seen by the CR in this process can be of the 
same order of magnitude as the grammage seen during its wandering through the Galaxy in a 
few tens of Myr (typically $\sim 10$~g/cm$^2$ at 1 GeV/nuc). The same mechanism which leads 
to the secondary galactic antimatter flux, leads as well to a secondary flux of the same 
particles via the interaction of CR particles with atmosphere (see \cite{HU03} for a 
quantitative comparison).

The CR induced \pb flux in the atmosphere has been measured recently at mountain level,
where it is expected to be of purely atmospheric origin \cite{BESS2770}. Light antinuclei 
have also been searched by the same experiment \cite{BESDBA}. These atmospheric secondaries 
will constitute a background for the CR flux of these particles in future embarked 
experiments. For balloon borne experiments this background, produced in the atmosphere above 
the detector, must be evaluated to correct the measurements \cite{BESS98}.
Besides, atmospheric particles can be trapped by the earth magnetic field and detected by 
satellite experiments. Although they could be separated in principle from the galactic flux 
on dynamic and kinematic grounds \cite{STORM}, a basic theoretical knowledge of this 
background is required.

As quoted in the introduction, the phenomenology of the particle production induced by CRs 
in the atmosphere has been thoroughly investigated recently with the purpose of accounting 
for the large amount of new data available from recent ballon and satellite measurements. 
The calculation of the antiproton flux was part of this effort \cite{HU03}. 
The latter have been recalculated in the present work with the tertiary (NAR) production 
taken into account, and extended to the case of A$>$1 antinuclei production on the same 
dynamical basis as for the galactic flux, as discussed below. The present results complement 
and update our previous calculations \cite{HU03}.
  \subsection {Antimatter flux}
\begin{figure}[hbt]          
\begin{center}
\includegraphics[width=\columnwidth,angle=0]{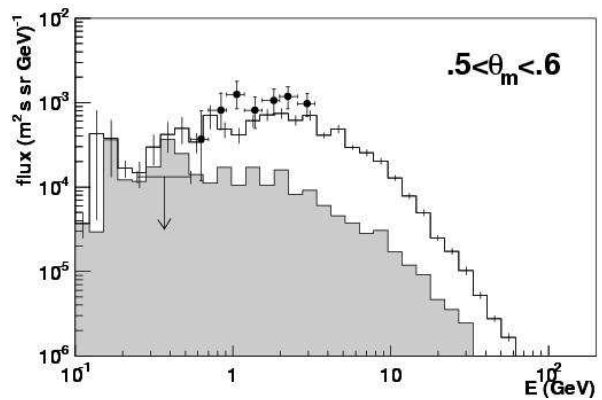}
\caption{\it\small Atmospheric antiproton flux at mountain altitude (2770 m) measured by the 
BESS collaboration \cite{BESS2770} (at the indicated geomagnetic latitude $\theta_m$), 
compared with the present calculations. The latter update the results presented in \cite{BA03}. 
The grey histogram corresponds to the NAR process, the white histogram showing the full 
calculations.
\label{PB2770}
}
\end{center}
\end{figure}
%
%
%
The atmospheric production of antimatter has been calculated along the same lines as in the 
previous calculations of the atmospheric particle flux by the authors \cite{HU03,BA03}. It 
consists of a 3D Monte-Carlo simulation program processing the propagation and interactions 
of charged particles in the earth environment \cite{DE01,LI02}. 
For antinuclei $A\leq$~4, the antimatter production cross sections used in the program have 
been calculated by means of the same standard coalescence model as described above, using 
the \pb production cross sections discussed in section~\ref{PBPROD}.
In the propagation process, the absorption cross sections for \pb$A$ collisions discussed in 
section~\ref{CSRPB} above were used, as well as the NAR contributions for the \pb and \db 
particles, discussed in section~\ref{NAR}.

Fig.~\ref{PB2770} shows the \pb flux at mountain altitude (2770~m), i.e., deep inside the 
atmosphere, measured by the BESS collaboration \cite{BESS2770}, compared to the present 
calculations. 
It can be observed that the ratio of the contribution of the tertiary \pb flux (grey 
histogram on the figure) to the parent secondary component, appears to be on the same scale as 
in the galactic flux, as expected. The overall flux appears to be somewhat ($\approx$20\%) 
smaller than the value reported in \cite{HU03}, due to the use of \pb$A$ absorption (total 
reaction) cross section (see above) smaller in the present calculations than in the previous 
report. The larger relative flux at small kinetic energies is due to the NAR process, taken 
into account in the present work while it was not in the previous one. The overall agreement
with the data, although slightly less good than in \cite{HU03}, is still fair however. 

\begin{figure}[hbtp]          
\begin{center}
\includegraphics[width=\columnwidth,angle=0]{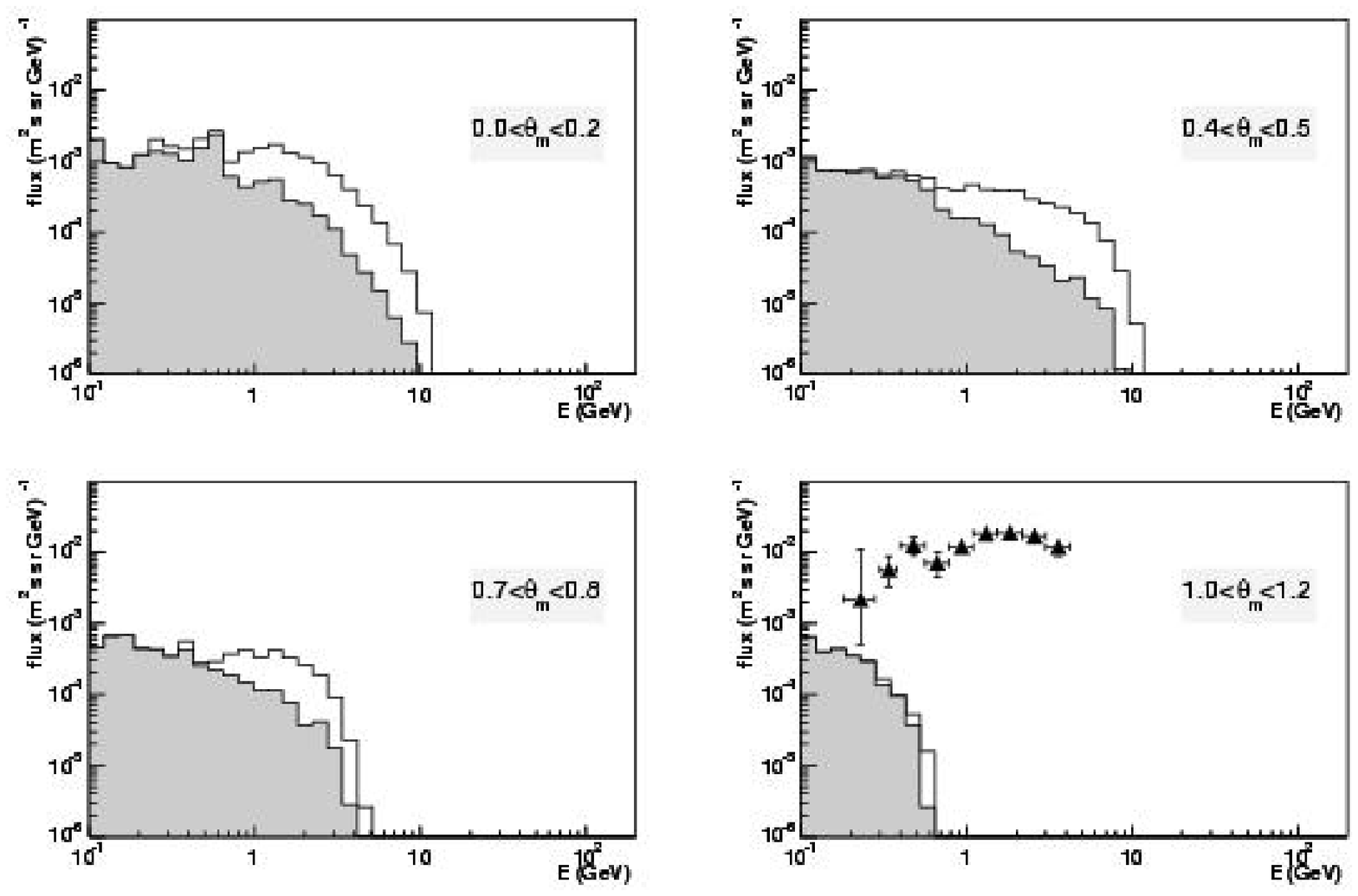}  \\
\includegraphics[width=\columnwidth,angle=0]{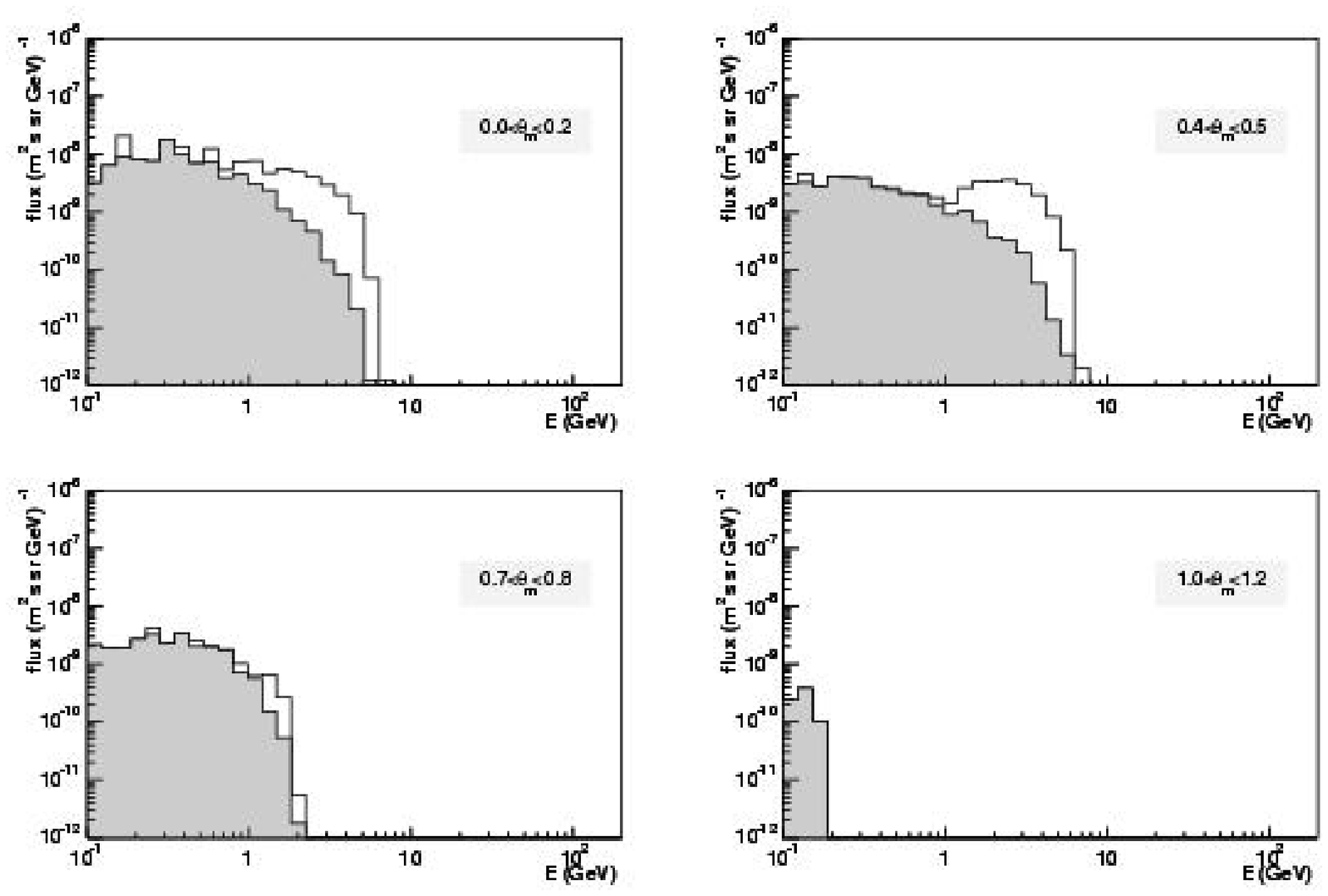}  \\
\includegraphics[width=\columnwidth,angle=0]{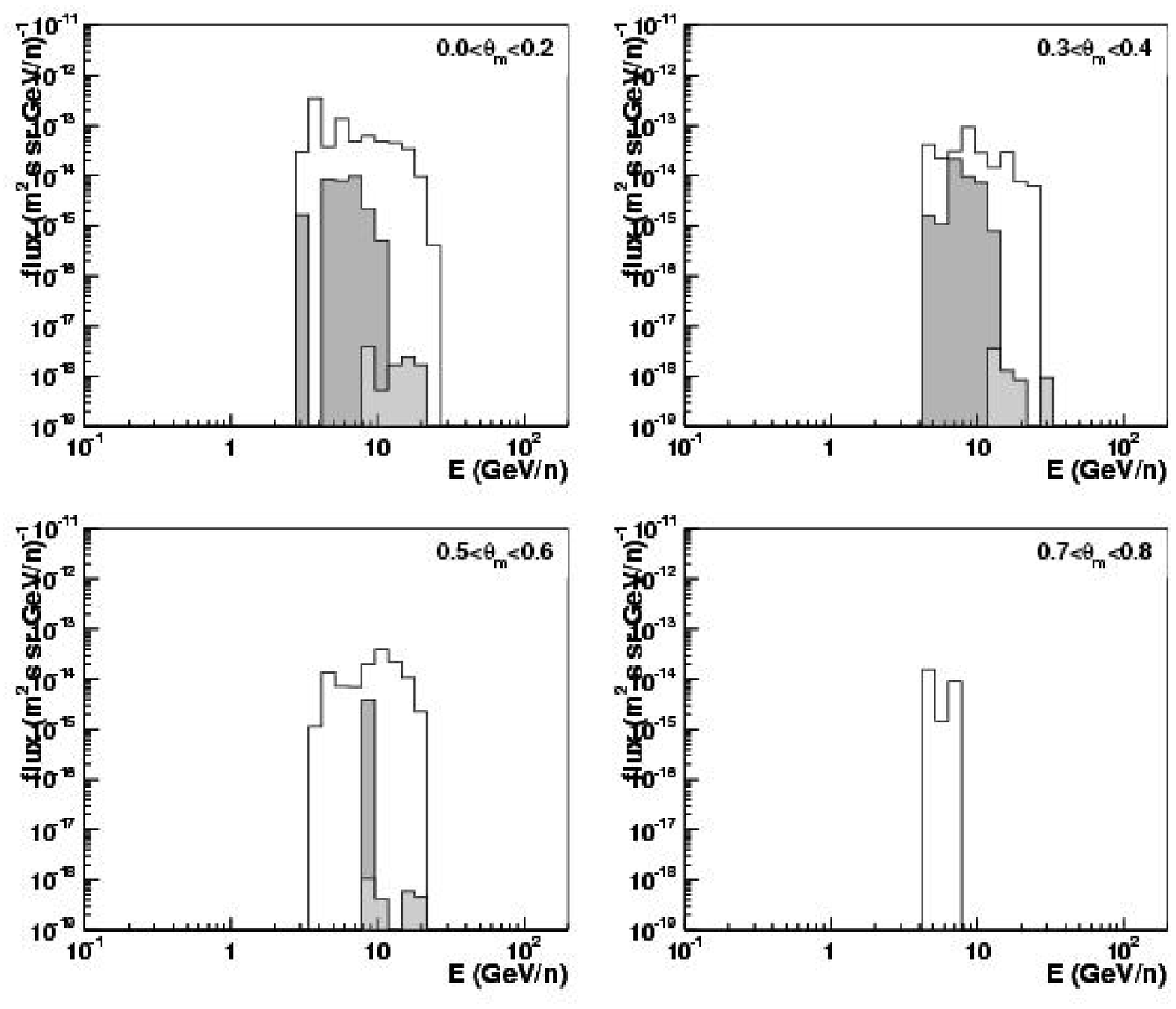} \\
\caption{\it\small Calculated fluxes of antimatter nuclei produced in the atmosphere, 
shown by groups of 4 panels, for \pb (top), \db (middle), with the full (secondary+tertiary, 
white histogram) and tertiary only (grey histogram) fluxes shown. The recent experimental 
\pb data from \cite{BESS99} for the galactic flux at TOA are also shown (full triangles).
The lower panel shows the mass 3 and 4 antinuclei fluxes, \tb (dark grey), \hetb (white), 
and \heqb (light grey) (bottom).
All calculations are for 400~km of altitude, and for the different bins in latitudes between 
equator and poles indicated on each individual panel ($\theta_m$ is the geomagnetic latitude 
angle in radian).
\label{PBDB3B400D}
}
\end{center}
\end{figure}

Fig.~\ref{PBDB3B400D} shows the downward flux of antimatter nuclei \pb, \db, and \tb, at 
400~km of altitude, induced by CR collisions with atmospheric nuclei, for a set of bins in 
latitude between equator and poles \cite{BA04}. Similar distributions are obtained for the 
upstream flux, showing that the dominant part of this flux is made of trapped particles 
\cite{DE01}.
A striking feature of the results is the very large proportion of the flux due to the
tertiary component (NAR process), while this fraction was much smaller both for the 
flux at ground level and for the galactic flux. This originates from the selectivity of the 
earth magnetic field which tends to trap more efficiently low energy particles, the NAR 
process decreasing the energy of the propagating particle then trapped by the field and 
confined in the belt. Note that in the polar region where the earth field is small, the 
whole flux is almost totally accounted for by the tertiary component. The same is true for 
the \db flux (intermediate panels).
The shape of the distributions of the secondary \pb and \db fluxes are also found 
significantly different from their galactic counterpart (see \cite{BA03a}). In particular, 
the atmospheric secondary particle spectra appear to have their maximum at different 
energies than galactic particles. 
All these features are due to the dynamics of the particles in the earth magnetic field which
may induce considerable distortion of the primary particle spectra \cite{BA04}.
\begin{figure}[hbtp]          
\begin{center}
\includegraphics[width=\columnwidth,angle=0]{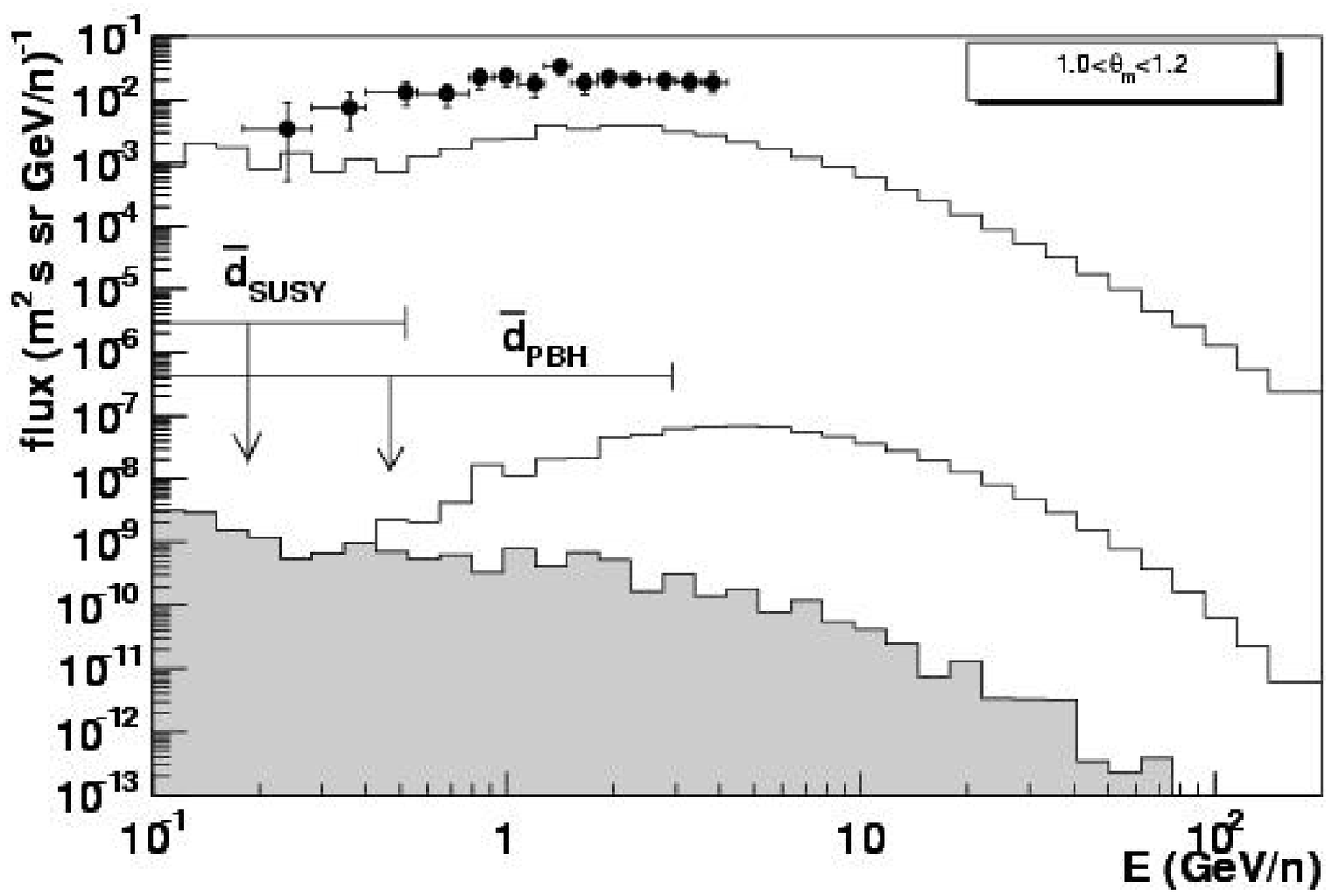} \\
\includegraphics[width=\columnwidth,angle=0]{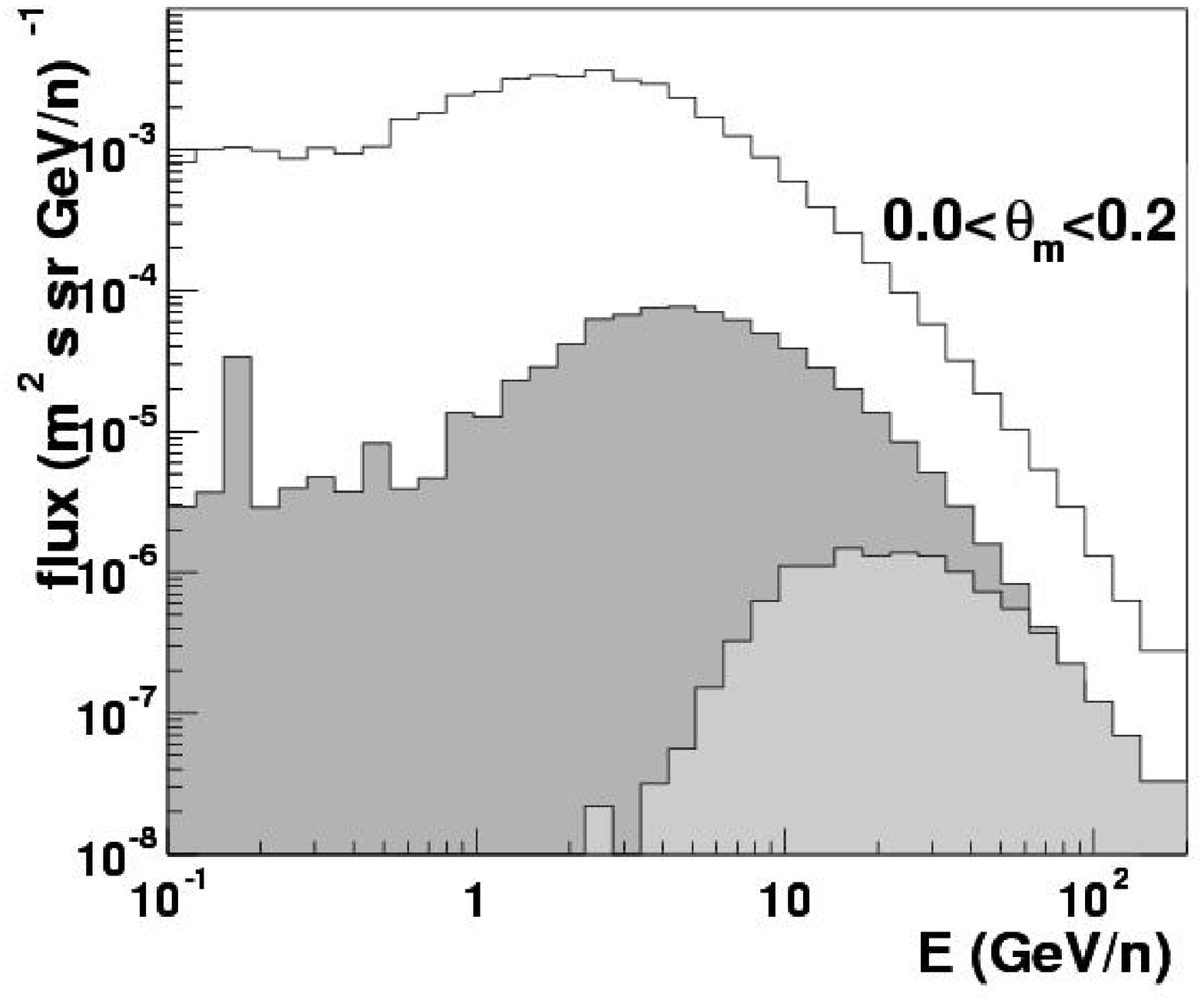}
\caption{\it\small Top: Calculated atmospheric \pb flux at 38~km of altitude (top histogram) 
compared with the recent BESS98 measurements \cite{BESS98} (which include both the 
galactic and the atmospheric components).
The lower two histograms correspond to the calculated atmospheric \db flux in the polar 
region, the gray histogram shows the tertiary component.
The horizontal line labelled $\bar{d}_{SUSY}$ corresponds to the limit for the \db flux 
supersymmetric dark matter annihilation, from \cite{DBDM}. 
Bottom: Differential atmospheric \pb, \db ($\times$5 10$^3$), \hetb($\times$10$^6$), fluxes 
in the polar region, at 38~km. For this latitude, the \hetb and \tb fluxes are identical 
(identical production cross sections). 
\label{DB38D}
}
\end{center}
\end{figure}
On Fig.~\ref{DB38D} (top panel), for the polar latitudes, the \pb flux compared with the TOA 
measurements by BESS shows that this atmospheric background is negligible at the considered 
satellite altitude for energies T$\gtrsim$~0.5~GeV. At lower energies however, the data and 
the atmospheric flux seem to converge towards close values.

Fig.~\ref{DB38D} lower panel shows the calculated spectral distributions for mass 3 and 
4 antinuclei. The typical structure has the same origin as for lighter antinuclei: the 
spectrum is limited by the rise of the particle production cross section driven by the 
production threshold (itself rising with the mass of the produced particle), on the lower 
side. The upper energy limit is set by the trapping conditions. 
For these fluxes the NAR contributions haven't been included. Dynamically, the same considerations 
as given in section~\ref{MASS34G} concerning the NAR process are valid as well for atmospheric particles.
However, as seen above, the expected NAR effect is much larger for atmospheric than for galactic 
fluxes. Including this process would somewhat weaken the high energy component and strongly 
populate the low energies (empty on the figure). But the induced change (increase of the overall 
flux), although significant at the considered order of magnitude, would not modify the conclusion, 
the overall flux being vanishingly small. 


  \section{Counting statistics for light antimatter nuclei at satellite altitude}
\begin{table}[h] 
\begin{center}
\begin{tabular}{cllll}
\hline\hline
Particles        &   \pb      & \db      &  \hetb       & \heqb        \\
\hline
Galactic rate    & 10$^6$     & 15       & 10$^{-3}$    &  10$^{-7}$    \\
Atmospheric rate & 5 10$^5$   & 3        & 10$^{-4}$    &  5 10$^{-9}$  \\
\hline\hline
\end{tabular}
\caption{\small 
Expected count rates for light antinuclei in the AMS experiment, for 3 years of effective counting
\cite{DU04,BA04}, as described in the text.
\label{INTGFL}}
\normalsize
\end{center}
\end{table}
The forthcoming search experiments for primordial antimatter will need some estimated values of 
the expected secondary galactic and atmospheric flux for the investigated particles, to be
confronted to the experimental results.
The secondary fluxes of antimatter particles calculated above have been energy integrated over 
the energy range 0.1-100~GeV per nucleon and over the AMS orbit taking into account the effects 
of the geomagnetic cutoff and using an approximate acceptance of 0.5~$m^{2}sr$ for the AMS 
spectrometer. The results are given in table~\ref{INTGFL}. The resulting counting statistics to 
be expected is of the order of 10$^6$, 20, 10$^{-3}$, and 10$^{-7}$ particles per year, for \pb, 
\db,\hetb, and \heqb, respectively. Clearly, the rates for mass 3 and 4 antinuclei are negligibly
small even at the scale of very small statistics. Note however that a \heqb candidate event would
have the largest probability of being a misidentified background \hetb particle.  
 
For the AMS experiment, a more accurate estimate of these numbers will be obtained with the 
processing of these fluxes in the AMS spectrometer simulation program, taking into account the 
identification capability of the instrument. The work is under way and the results will be 
reported later.
\section{Summary}\label{CONC}
This work has provided new calculations of the light antinuclei \pb,\db,\hetb,\heqb, fluxes 
produced by CR collisions on the ISM in the Galaxy, and the first calculations of the fluxes 
produced by CRs in the atmosphere for the same particles. For the \db flux, a new production 
channel $\bar{p}p\rightarrow\bar{d}X$ has been included in the calculations and was shown to 
contribute significantly to the total flux.
The cross sections involved in the calculations were based to a large extent on existing 
accelerator data. The results are found significantly different from previous calculations. 
The calculated fluxes above 0.5-1~GeV/n are estimated to be reliable within a factor of 2 for 
the \db flux, and within one order of magnitude for mass 3 antinuclei. The calculated flux 
in the low energy range, below 0.5~GeV/n, suffers larger uncertainties mainly because of the 
lack of experimental data. The calculated \db flux appears to be marginally detectable by 
the AMS experiment mainly because it will have to be discriminated against a large 
background of \pbs and electrons.

A further step will consist of using the calculated \db flux as an input to the AMS02 
spectrometer simulation together with the other particles with the same electric charge 
(\pb, electrons) with their respective galactic fluxes, and discriminating the \dbs from the 
other particles by means of the spectrometer instruments, in order to evaluate the capacity 
of the experiment to achieve the measurement of this flux.

Acknowledgements: The authors are indebted to P. Salati and R. Taillet for several 
enlightening discussions on the subject. The diffusion model calculations are
based on a code originally written by P.~Salati.


\newpage
\appendix
\section*{Appendix: Inelasticity in composite particle scattering}
Let us consider the system for incident deuterons rather than antideuterons, first. The result
will be easily extended to \db later. Qualitatively the effect of the small dissociation
energy of the deuteron, or of any nucleus or other quantal system with small binding energy,
is to make the collision highly diffractive because of the strong absorption induced by the
large dissociation cross section of the deuteron. Because of this larger collision opacity,
the inelastic collisions on nuclei will be more peripheral than they would be for an incident
nucleon for example. For the same reason the angular distribution will be more forward peaked,
partly because of the ``fragility'' of the projectile which limits the range of
momentum transfer to low values, as explained below. But this does not imply that they would
have a smaller cross section. In fact, it can be shown that to the contrary, the inelastic
collision cross section induced by a composite nuclear projectile is expected to be larger
than for incident nucleons. This is also supported by the existing data.

In the case of $dp$ collisions on a nucleon target however, the system is rather transparent
and the diffractive effect can only be loosely referred to, the forward peaking of the
cross section being induced mainly by the deuteron form factor. As discussed below, the
latter is broader in the momentum space and the resulting angular distribution more forward
peaked, which means that only low momentum transfers are allowed in the scattering process.

All the above arguments apply to the case of incident antideuterons which collisions are made
even more diffractive than for incident deuterons by the stronger absorption induced by the
annihilation channels. The following arguments applies equally well to both cases.

In the type of collision considered here, the incident \db excites the ISM nucleon into a
resonance-like state with mass M$_X$. The process is dominantly diffractive, characterized by
small momentum transfer and peripheral character. The hit nucleon can as well be bound inside
a (ISM) nucleus.

The transition amplitude $F(t)$ ($t$ 4-momentum transfer) for such a reaction can be most
conveniently described for the present pedagogical purpose, in the single scattering limit
of the Glauber multiple scattering approach for the collisions of composite systems
\cite{FR78,BU90}, namely the impulse approximation, in terms of the form factors of the
interacting systems $G_p(t)$, $G_t(t)$ and of the interaction term $v(t)$.
The full Glauber amplitude can be written as :
\begin{equation}
F(t)=\int d^2b e^{i\vec{q}\cdot\vec{b}} <\Psi_p\phi^*_t|\Gamma(b,s_p,s_t)|\Psi_p\phi_t>
\label{GLAU}
\end{equation}
where $\Psi_p$, $\phi_t(\phi^*_t)$ are the projectile and target ground (excited) states
wave functions respectively,  $b$, $s_p$, and $s_t$, the impact parameters variables,  
and $\Gamma(b,s_p,s_t)=1-e^{i\sum\chi_j}$ 
the (Glauber) profile function of the collision system, with $\chi_j$ being the individual
phase shift functions of the elementary constituents \cite{AL81}, namely nucleons for a 
nuclear system \cite{FR78}, constituent quarks for a subnucleon system \cite{HA68}, or both 
\cite{FO87}. This later relation is the foundation of the Glauber approximation and leads to 
the well known multiple scattering series of the scattering amplitude.

A classical illustration of the correctness of the Glauber approach is precisely the
application to the total deuteron cross section on proton, for which the above series for
the case of elastic scattering using the optical theorem, leads to \cite{FR66}:
$$\sigma_{tot}(dp)\approx \sigma(pp) + \sigma(np) - (2nd\_order\_shadowing\_term) $$

In the single scattering approximation, i.e., the Plane Wave Impulse Approximation (PWIA)
(or Chou-Yang model for hadron-collisions), the elastic scattering amplitude takes the very
simple form:
\begin{equation}
F(t)\sim A_pA_tG_p(t)G_t(t)v(t)
\label{IMPA}
\end{equation}
Where the numbers of constituents of the two systems $A_p$ and $A_t$ originate from the
normalization of the ground state wave function \cite{FR78,BU90}.
From this relation, it can be shown straightforwardly by simple application of the optical
theorem, that the total reaction cross section is proportional to the product of the number
of constituents times the elementary total cross section between constituents
$\sigma_{tot}=A_pA_t\sigma_0$. This is the additive quark model for hadron-hadron collisions
\cite{LI66}.

The derivation for the inelastic $Dp\rightarrow DX$ cross section goes through identical
steps with the only difference that the proton final state is an excited state rather than
the ground state, and the product of the initial and final proton wave functions in the
relation above which describes the ground state density in the case of elastic scattering,
then becomes a transition density describing the excited nucleon state
$\delta\rho(r)$ instead of the ground state density $\rho(r)$.

A similar relation applies for inelastic scattering. In that case, one (or both) of the form
factors involved describe the excited state, namely: \\
$F(t)\sim A_pA_tG_p(t)v(t)\delta_t(t)$

Therefore the expected inelastic $Ap\rightarrow AX$ cross section, A being a composite system
such as the deuteron $d$, is expected to be larger than its counter part in nucleon-nucleon
scattering $pp\rightarrow pX$.

This result is confirmed by the experimental data from \cite{GO79} (see also \cite{AL81}) where
a larger cross section for  $dn\rightarrow d(p\pi^-)$ than for $pn\rightarrow p(p\pi^-)$ was 
measured (see table 2 and fig.6 in \cite{GO79}, see also \cite{BA88}).
%

%
\end{document}